\documentclass[3p, twocolumn]{elsarticle}
\usepackage{numcompress}
\biboptions{sort&compress}

\usepackage{hyperref}
\usepackage{afterpage}
\usepackage{xcolor}
\hypersetup{colorlinks=true, linkcolor=[rgb]{0,0,0.75}, citecolor=[rgb]{0,0.75,0}}
\usepackage{booktabs}

\journal{Sustainable Energy Grids and Networks (SEGAN). Accepted for publication.}

\usepackage[utf8]{inputenc}
\usepackage[cmex10]{amsmath}
\allowdisplaybreaks
\usepackage{siunitx}
\usepackage{array}
\usepackage{amssymb}

\usepackage{tikz}
\usetikzlibrary{decorations.pathreplacing}
\usetikzlibrary{arrows,shapes,positioning}
\usetikzlibrary{patterns}
\usepackage{pgfplots}
\pgfplotsset{compat=newest}
\pgfplotsset{plot coordinates/math parser=false}
\newlength\figureheight
\newlength\matlabfigurewidth
\newcommand{\ie}{{i.e.\ }}
\newcommand{\eg}{{e.g.\ }}
\newcommand{\citefig}[1]{Fig.~\ref{#1}}
\newcommand{\citeeq}[1]{{Eq.~\eqref{#1}}}

\newcommand{\flexhouse}{{{Power~Flexhouse}}}

\newcommand{\citetable}[1]{{Table~\ref{#1}}}
\newcommand{\citesec}[1]{{Section~\ref{#1}}}

\newcommand{\COP}{{\mbox{COP}}}

\newcommand{\argmin}[1]{\underset{#1}{\operatorname{arg}\,\operatorname{min}}\;}

\newcolumntype{L}[1]{>{\raggedright\let\newline\\\arraybackslash\hspace{0pt}}m{#1}}
\newcolumntype{C}[1]{>{\centering\let\newline\\\arraybackslash\hspace{0pt}}m{#1}}
\newcolumntype{R}[1]{>{\raggedleft\let\newline\\\arraybackslash\hspace{0pt}}m{#1}}

\usepackage{xstring}
\newcommand{\paper}[1]{{{\ignorespaces
  Paper~\textbf{\IfEqCase{#1}{%
    {pes}{[\ref{paper:pes}]}%
    {isgt}{[\ref{paper:isgt}]}%
    {freezer}{[\ref{paper:freezer}]}%
    {repl}{[\ref{paper:repl}]}%
    {upec}{[\ref{paper:upec}]}%
    {cdc}{[\ref{paper:cdc}]}%
    {gm}{[\ref{paper:gm}]}}[\PackageError{paper}{Paper not defined: #1}{}]%
}}}}

\begin{document}


\title{Grey-box Modelling of a Household Refrigeration Unit Using Time Series Data in Application to Demand Side Management}

\author[dtuelektro]{Fabrizio Sossan\corref{mycorrespondingauthor}}
\ead{fabrizio.sossan@epfl.ch}
\author[dtuelektro]{Venkatachalam Lakshmanan}
\author[dtuelektro]{Giuseppe Tommaso Costanzo}
\author[dtuelektro]{Mattia Marinelli}
\author[dtuelektro]{Philip~J.~Douglass}
\author[dtuelektro]{Henrik Bindner}
\cortext[mycorrespondingauthor]{Corresponding author}
\address[dtuelektro]{DTU Elektro, Frederiksborgvej 399, 4000, Roskilde, Denmark}

\begin{frontmatter}

\begin{abstract}
This paper describes the application of stochastic grey-box modeling to identify electrical power consumption-to-temperature models of a domestic freezer using experimental measurements.
The models are formulated using stochastic differential equations (SDEs), estimated by maximum likelihood estimation (MLE), validated through the model residuals analysis and cross-validated to detect model over-fitting.
A nonlinear model based on the reversed Carnot cycle is also presented and included in the modeling performance analysis.
As an application of the models, we apply model predictive control (MPC) to shift the electricity consumption of a freezer in demand response experiments, thereby addressing the model selection problem also from the application point of view and showing in an experimental context the ability of MPC to exploit the freezer as a demand side resource (DSR).

%
%
%
%
%

\end{abstract}

\begin{keyword}
Power demand \sep Modelling \sep Refrigerators \sep Smart grids \sep Demand Response.
\end{keyword}

\end{frontmatter}


\section{Introduction}
Household refrigerators account for a noticeable share of the total residential electricity demand (\eg 7\% in the US \cite{energyoutlook2012}) and are gaining attention in the context of demand side management (DSM) \cite{Xydis201341, 6558529, 6598997}.
Validated mathematical models and procedures for on-line system identification of refrigeration units are of importance for assessing their energy efficiency, predicting their power consumption and in application to intelligent energy management strategies to support power system operation, such as model predictive control (MPC) \cite{Hovgaard2012105, Hovgaard2013InJC}. 
In the first part of this paper, we describe the application of a state-of-the-art grey-box modeling methodology to identify power consumption-to-temperature prediction models using experimental measurements from a conventional domestic freezer. 
The modeling effort aims to identify the model structure and parameters of the physical processes associated with the operation of the freezer, \ie\ heat transfer and refrigeration cycle coefficient of performance (\COP). Consumer behavior modeling is not considered at this stage. 
In the second part of the paper, the proposed models are used to implement a MPC strategy in order to experimentally achieve a shift in the energy consumption of the freezer. 
The contribution of this paper is twofold. First, novel validated grey-box models for a domestic freezer are proposed. Existing models in the literature were mainly developed using first principle approaches (see for example \cite{hermes2009prediction, hermes2009assessment, 5930429}). So-called white-box models in application to demand side management do not allow to achieve any degree of differentiation when dealing with different units. This property does not make them suitable for future smart grid scenarios, where demand response built up from the contribution of heterogeneous populations of demand side resources (DSRs) is expected to play a central role in assuring reliable power system operation.
On the contrary, grey-box models are adaptive by nature since they are estimated from measurements and potentially allow for tracking system changes on-line by reiterating the estimation procedure.
A set of linear grey-box models for a domestic fridge was proposed in \cite{costanzo2013grey}. In this case, we extend the already existing literature by considering a domestic freezer and, especially, by including in the modeling performance assessment a nonlinear grey-box model based on the reversed Carnot cycle. 
Second, by implementing the consumption shift experiments in a real operating environment, we address the model selection problem from the specific perspective of the application, that is finally among the most relevant problems for DSM. Overall, we therefore provide a global assessment of grey-box modeling for refrigeration units analyzing both the pure modeling performance and application.
The structure of the paper is as follows. \citesec{sec:expSetup} describes the experimental setup adopted for the model identification and consumption shift experiments. \citesec{sec:greybox} describes the grey-box modeling framework adopted to identify the freezer models, which are therefore presented in \citesec{sec:models}. In \citesec{sec:modelsperformance}, we perform an empirical evaluation of the thermal properties of the considered freezer with the objective of supporting grey-box modeling results. 
In \citesec{sec:models}, the models prediction performance are further assessed using validation data sets.
Finally, in \citesec{sec:application} a number of the proposed model are used in an MPC experiments with the objective of shifting the consumption of a freezer in the context of intelligent energy strategies for demand response applications.


\section{Experimental setup}\label{sec:expSetup}
The experimental setup consists of a freezer equipped with temperature sensors, a power consumption measurements board and an external relay.  
The objective of the experiment is to collect the measurements for identifying the freezer models and to perform the energy shift experiments.
The instrumented freezer (shown in \citefig{fig:expsetup}) is a Bosch GSN40A21\footnote{The freezer belongs to \flexhouse, an experimental facility of DTU Elektro for testing demand side management strategies for smart grid applications.}, a commercially available domestic unit with \SI{333}{L} capacity and a single-phase compressor. During experiments, the freezer was empty and with closed door. 
Temperatures are measured using 10k NTC thermistors, which can measure temperatures in the range -30 to \si{80\celsius} with an accuracy of \SI{\pm0.2}{\celsius} at \si{25\celsius} and have a fast measurements response. Thermistors are connected to a 12bit ADC through a resistance-to-voltage transducer. A total of 3 thermistors are used, 2 for measuring the freezer interior temperature at different heights and one for the room temperature.
The freezer power consumption is measured using a DEIF MIC-2. This accuracy class 0.5 instrument is able to measure voltages and currents up to \SI{400}{V}/\SI{5}{A} of magnitude on a three phases bus, although only one phase is used for this application. 
The controllable power plug determines the state (on-off) of the freezer. In order to override the internal action of the freezer thermostat, the thermostatic set-point is set to the lowest value, and the temperature of the freezer is allowed to vary only above this threshold. In this way, the activation of the freezer compressor depends only on the state of the external relay. 
All the sensors and instruments are connected to a PC and are accessed by a JAVA software application. Measurements and actuations are sampled at 1~second.

\begin{figure}[!ht]
\centering
\scriptsize
\includegraphics[width=1\columnwidth]{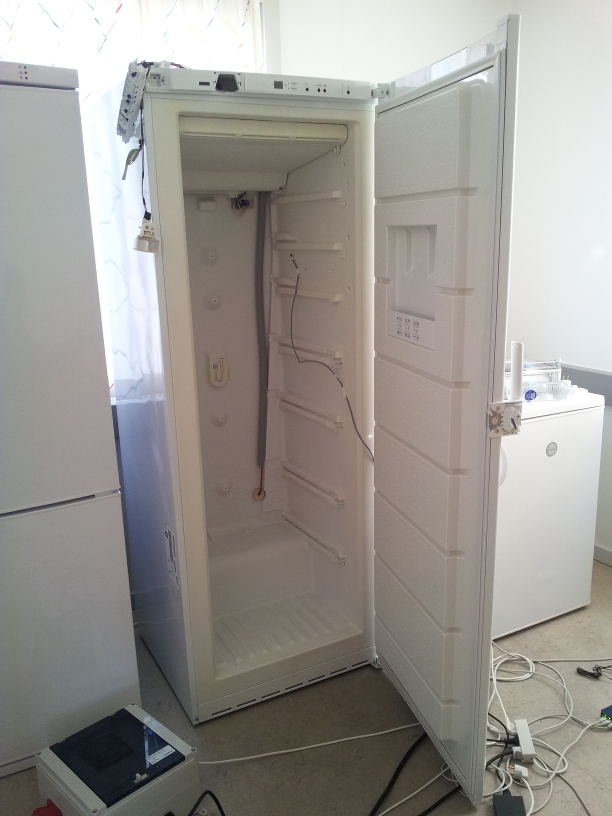}
\caption{The 333 liters domestic freezer used for the identification and MPC experiments.}\label{fig:expsetup}
\end{figure}

\section{The grey-box modeling methodology}\label{sec:greybox}
Grey-box modeling is a framework to identify and validate a mathematical model of a system incorporating its physical knowledge together with measurements from a real device. It consists of a number of steps, which are explained in the following.

\subsection{Experimental design} A control signal commonly used for model identification is the PRBS (pseudo binary random signal), an on-off signal with fixed period and uniformly distributed random duty cycle that is able to excite the system to model in a wide range of frequencies. 
For identifying the freezer thermal models, a PRBS was used to set the state of the controllable power plug.
To avoid damaging the freezer compressor, the PRBS cycles with on-to-off transitions shorter than 30~seconds were disregarded. By doing this, it was not possible to observe very short transients, which however are of limited interest since we target to capture the dominant system dynamics.
\citefig{fig:dataestimation} shows the set of measurements used for the model identification. In the upper panel plot, it is evident the effect of the PRBS on the freezer power consumption, that is characterized by activation cycles of random length. 
The time series are of appropriate duration for the purpose of thermal models identification as they are considerably longer than the slowest time constant of the system ($\approx 4~$hours\footnote{Approximatively estimated prior the model identification as $1/5$ of the zero-input temperature transient duration.}).
Two additional sets of measurements were collected to assess the models prediction performance (\citesec{sec:modelsperformance}).

\begin{figure}[!ht]
\centering
\scriptsize
\input{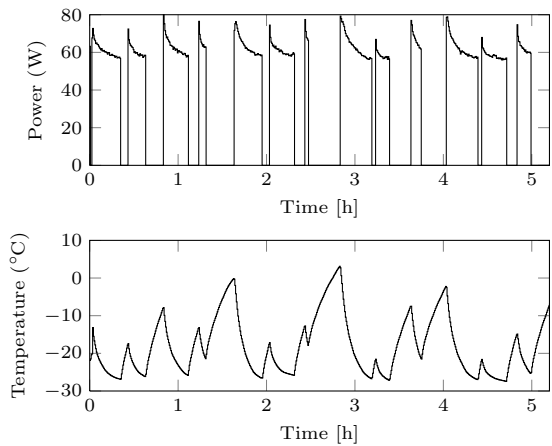}
\caption{The measurements used to identify the freezer thermal models: the freezer power consumption under PRBS regime (upper panel) and temperature (lower panel).}\label{fig:dataestimation}
\end{figure}

\subsection{Measurements post-processing}
The measured physical quantities are the freezer and room temperatures and freezer total power consumption. 
Two freezer temperature sensors are available, and their readings are averaged in order to obtain a single temperature measurements time series.
Power consumption measurements were treated to remove the components not playing a role in the refrigeration cycle since they can interfere with the estimation of the model parameters. First, the power absorption of auxiliary components (like the control logic and digital display), considered constant, was removed by subtracting. 
Second, the power consumption spikes due to the inrush current of the induction motor driving the compressor were filtered out.
The spikes usually extinguish in a few seconds and are disregarded because they represent the energy spent to accelerate the compressor and have a negligible impact on the thermodynamic transformation operated by the freezer.

\subsection{Model Formulation} \label{methods:model_formulation}
A set of mathematical relationships to describe the physical process to model is formulated. In general, refrigeration units, like freezers and fridges, implement a thermodynamic cycle to accomplish heat transfer from a cold to a hot reservoir by supplying mechanical work to the system. 
The freezer models are derived using the thermal equivalent circuit (TEC) approach, that considers the temperatures and heat fluxes of a thermal system as voltages and currents of an electric circuit. 
The proposed models as well as the basic thermodynamic concepts adopted to derive them are detailed in the following section.
The models are formulated using stochastic differential equations (SDEs) and continuous time state space representation. In the linear case, the models are:
\begin{align}
& d\boldsymbol{x}_t = A(\boldsymbol{\theta})\boldsymbol{x}_tdt + B(\boldsymbol{\theta}) \boldsymbol{u}_t dt + W(\boldsymbol{\theta}) d\boldsymbol{\omega} \label{eq:ssd:1} \\
& T_k = C \boldsymbol{x}_k + v(\boldsymbol{\theta}) e_k \label{eq:ssd:2}
\end{align}
with $\boldsymbol{x} \in \mathbb{R}^{n}$ as the state vector, $\boldsymbol{u} \in \mathbb{R}^{p}$ input vector, $A \in \mathbb{R}^{n\times n}$ system matrix, $B \in \mathbb{R}^{n\times p}$ input matrix, $\boldsymbol{\theta} \in \mathbb{R}^m$ a vector of $m$ model parameters, $W \in \mathbb{R}^{n\times n}$ process noise matrix (diagonal), $T$ the estimated freezer temperature, $C \in \mathbb{R}^{1\times n}$ output vector, $V \in \mathbb{R}$ measurement noise, and where the subscripts $t$ and $k$ respectively denote continuous and discrete time quantities. 
The vector $\boldsymbol{\omega}$ is a standard $n$-dimension Wiener process, \ie\ a continuous time stochastic process with independent normally distributed increment, while $e_k$ is normal Gaussian noise. \citeeq{eq:ssd:1} describes the stochastic evolution of the system, while \eqref{eq:ssd:2} is the observation equation and links the model state to the estimated freezer temperature.  


\subsection{Maximum likelihood estimation of the parameters} \label{methods:par_estimation}
The parameters of the candidate model are estimated utilizing CTSM, a system identification software library for R that implements maximum likelihood estimation (MLE) \cite{juhl2013ctsm}. In the following, a general insight on the method is given. For a comprehensive description into the algorithm, the reader is referred to \cite{4336b714bb4b4ad8869e62d0bf115bf5}, and to \cite{CTSMWall} for an alternative modeling application. Given a model in the form as in \eqref{eq:ssd:1}-\eqref{eq:ssd:2}, the objective of MLE is to determine the unknown parameters $\boldsymbol{\theta}$ by maximizing the likelihood of the model getting the observed training data set.
Given a time series of the output variable (such as the freezer temperature) denoted as $\mathcal{Y}_N = \{y_0, \dots, y_N\}$, the model likelihood function is derived starting from the joint probability density:
\begin{align}
L(\boldsymbol{\theta}, \mathcal{Y}_N) &= p\left(\mathcal{Y}_N | \boldsymbol{\theta}\right),
\end{align}
that can be reformulated as a product of conditional probability densities, like
\begin{align}
 L(\boldsymbol{\theta}, \mathcal{Y}_N) =\left(\prod_{k=1}^Np(y_k|\mathcal{Y}_{k-1}, \boldsymbol{\theta})\right)p(y_0|\boldsymbol{\theta}),\label{eq:likelihoodf}
\end{align}
by applying the chain rule. The output of the stochastic model is completely characterized in terms of mean and variance, which are respectively given as:
\begin{align}
\begin{aligned}
 \hat{y}_{k|k-1} &= E\big[{y}_{k}|\mathcal{Y}_{k-1}, \boldsymbol\theta\big]\\
 R_{k|k-1} &= V\big[{y}_{k}|\mathcal{Y}_{k-1}, \boldsymbol\theta\big].
\end{aligned}
\end{align}
The one step ahead prediction errors of the model, or model residuals, are defined as the following scalar sequence:
\begin{align}
 \epsilon_k = y_k - \hat{y}_{k|k-1},\;\; k=1,\dots,N \label{eq:ctsm:l5},
\end{align}
and the model likelihood function can be reformulated as
\begin{align}
L(\boldsymbol{\theta}, \mathcal{Y}_N) = \left( \prod_{j=1}^k \dfrac{\exp\left(\frac{1}{2}\epsilon_j^2 R_{j|j-1}^{-1}\right)}
{\sqrt{2\pi R_{j|j-1}}} \right)p(y_0|\boldsymbol{\theta}) \label{likelihoodfunction},
\end{align}
where the formulation of the zero-mean univariate Gaussian distribution has been used.
The model parameters are finally found by CTSM through minimizing the negative logarithm of the likelihood function:
\begin{align}
 \boldsymbol{\theta}^o = \argmin{\boldsymbol\theta  \in \Theta} - \mathcal{L}(\boldsymbol\theta, \mathcal{Y}_N)\label{eq:loglikelihood}.
\end{align}
where
\begin{align}
 \mathcal{L}(\boldsymbol\theta, \mathcal{Y}_N) = \text{ln}(L(\boldsymbol\theta, \mathcal{Y}_N)) \label{eq:loglikelihoodl}
\end{align}
is called model log-likelihood.

\subsection{Model validation through the residuals analysis} \label{methods:model_validation}
This phase consists in evaluating if the freshly identified model was able to capture all the time dynamics contained in the measurements. 
The model validation is carried out with the residuals analysis, that consists in evaluating any correlation in time of the model residuals (defined in \eqref{eq:ctsm:l5}): if a model is acceptable, the model residuals should not show any autocorrelation or, in other words, the model explains all the dynamics in the measurements. A practical example of model validation is presented in the next section.

\subsection{Model extension and cross validation} \label{methods:model_expansion}
If the validation results are not satisfactory, an alternative model should be formulated by, for example, increasing the order of the previous model or adopting an alternative mathematical description of the process.
The extended model parameters should be estimated and validated by reiterating the steps described in the last two paragraphs. If the model extension consisted in augmenting the model order, the following procedure is applied to test the hypothesis that the model extension is valid \cite{madsen2011introduction}. Given the models A and B respectively with parameters $\boldsymbol\theta \in \Omega_A \subset \mathbb{R}^m$ and $\boldsymbol\theta \in \Omega_B \subset \mathbb{R}^n$ ($m$ and $n$ number of parameters, $m < n$), and with $ \Omega_A \subset \Omega_B$ (in other words, the simpler model is contained in the second), we test the null hypothesis $\mathcal{H}_0:\boldsymbol\theta \in \Omega_A$ against the alternative $\mathcal{H}_1:\boldsymbol\theta \in \Omega_B$. 
The deviance (or likelihood-ratio) for the current model extension is defined as:
\begin{align}
 D = 2\left(\mathcal{L}(\boldsymbol\theta \in \Omega_B, \mathcal{Y}_N) - \mathcal{L}(\boldsymbol\theta \in \Omega_A, \mathcal{Y}_N)\right) \label{eq:deviance}
\end{align}
where $\mathcal{L}$ is the model log-likelihood as defined in \eqref{eq:loglikelihoodl}. The Wilk's theorem states that, under the null hypothesis $\mathcal{H}_0$, the deviance \eqref{eq:deviance} converges in law to a chi-squared distribution with $k-m$ degree of freedom (denoted by $\chi^2(k-m)$-distribution).
Plainly from \eqref{eq:deviance}, a large deviance is in favor of the extended model. We use Wilk's theorem to define an upper threshold above which the deviance is not likely to be drawn from the $\chi^2(k-m)$-distribution, indeed giving evidence against the null hypothesis. More specifically, $\mathcal{H}_0$ is rejected at $\alpha$ confidence level (usually 95\%) if the following condition is verified:
\begin{align}
 D > \Phi_{k-m}^{-1}(\alpha),
\end{align}
where $\Phi_{k-m}^{-1}$ is the inverse CDF of the $\chi^2(k-m)$-distribution.
Equivalently, we can look at the probability of a new realization (of known distribution under $\mathcal{H}_0$) being larger than the current $D$, \ie\
\begin{align}
 p &= \text{Pr}(x > D),\ \ \ \ x \in \chi^2(k-m)
\end{align}
that can be reformulated as:
\begin{align}
\begin{aligned}
 p &= 1 - \text{Pr}(x < D) \\
   & = 1 - \Phi_{k-m}(D).
\end{aligned}\label{eq:pvalue}
\end{align}
where $\Phi_{k-m}(D)$ is the CDF of the $\chi^2(k-m)$-distribution calculated in $D$ and can be determined, for example, using the Matlab command $\textrm{chi2cdf}(D, k-m)$.
In \eqref{eq:pvalue}, $p$ is said to be the p-value. If the p-value is smaller than $1-\alpha$ (5\%), the null hypothesis $\mathcal{H}_0$ is rejected and the model extension is considered valid. This procedure acts as a stop condition of the model extension process. In fact, an extended model usually performs better than the original because have a larger number of parameters, but it should be disregarded  if the null hypothesis is failed to reject. A practical example of the application of this method will be provided in \citesec{sec:models} when cross validating the models.


\section{Freezer thermal models}\label{sec:models}
In this section, the identified freezer models are presented in increasing order of complexity. As explained in the previous section, models are derived using the TEC analogy, formulated using stochastic state space representation, estimated by MLE, validated through the model residuals analysis and, in case of model extensions, cross-validated to detect model overfitting. For each model, the mathematical formulation is presented, while the estimated parameters of all the models are summarized in Appendix A.

\subsection{Model~A (linear first order)}
\subsubsection{Model formulation}
\citefig{fig:modelA:circuit} shows the TEC of Model~A. The current $Q$ represents the heat that is extracted from the freezer interior. In this and following linear models, $Q$ is modelled as a function of the measured freezer power consumption $P$ according to the following relationship:
\begin{align}
Q = P \cdot \COP  \label{eq:modelA:Q},
\end{align}
where \COP\ denotes the coefficient of performance of the refrigeration cycle, a dimensionless quantity that expresses the amount of heat transferred from the cold to the hot heat reservoir per unit of mechanical work supplied to the cycle. It is worth to note that, in the expression above, the mechanical work is approximated with the electrical power consumption. Therefore in \eqref{eq:modelA:Q}, the \COP\ coefficient should be regarded as a lumped gain that describes the coefficient of performance while accounting for power conversion losses. 
The capacitor $C_a$ represents the global heat capacity of the freezer thermal mass (\ie\ air content, heat exchanger and freezer envelope) and $V_a$ corresponds to the measured freezer temperature. The electrical current flowing through the resistance $R_w$ represents the heat losses through the freezer envelope towards the environment, that is at the room temperature $T_r$. 
The Kirchoff's voltage law (KVL) applied to the circuit in \citefig{fig:modelA:circuit} constitutes the deterministic skeleton of Model~A. With reference to the stochastic state space representation \eqref{eq:ssd:1}-\eqref{eq:ssd:2} introduced in \citesec{methods:model_formulation}, the model state and input vector are as:
\begin{align}
\begin{aligned}
\boldsymbol{x}^T &= V_a  \\
\boldsymbol{u}^T &= \begin{bmatrix} T_r & P \end{bmatrix},
\end{aligned}
\end{align}
where $T_r$ and $P$ are the room temperature and freezer power consumption from measurements.
The stochastic system matrices are:
\begin{align}
A &= -\dfrac{1}{C_a R_w} \\
B &= \begin{bmatrix} \frac{1}{C_a R_w} & -\frac{\COP}{C_a} \end{bmatrix} \\
W &= w \\
C &= 1
\end{align}
where $C_a$, $R_w$, \COP\ and $w$ are the model parameters to be estimated. Their MLE-estimated values are in \citetable{tab:modelA:pars}.

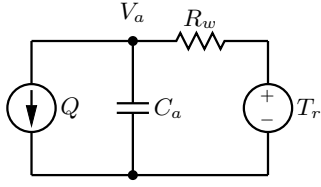
\begin{figure}[ht]
\centering
\small
\begin{tikzpicture}[scale=2.54]
\ifx\dpiclw\undefined\newdimen\dpiclw\fi
\global\def\dpicdraw{\draw[line width=\dpiclw]}
\global\def\dpicstop{;}
\dpiclw=0.8bp
\dpiclw=0.8bp
\dpiclw=0.85bp
\dpicdraw (0,0.7)
 --(0,0.475)\dpicstop
\dpicdraw (0,0.35) circle (0.049213in)\dpicstop
\dpicdraw (0,0.44375)
 --(0,0.3625)\dpicstop
\filldraw[line width=0bp](-0.026562,0.3625)
 --(0,0.25625)
 --(0.026563,0.3625) --cycle
\dpicstop
\dpicdraw (0,0.225)
 --(0,0)\dpicstop
\draw (0.125,0.35) node[right=-1.5bp]{$ Q$};
\dpicdraw (0,0.7)
 --(0.525,0.7)\dpicstop
\draw (0.525,0.85) node{$V_a$};
\dpicdraw[fill=black](0.525,0.7) circle (0.007874in)\dpicstop
\dpicdraw (0.525,0.7)
 --(0.525,0.375)\dpicstop
\dpicdraw (0.441667,0.375)
 --(0.608333,0.375)\dpicstop
\dpicdraw (0.441667,0.325)
 --(0.608333,0.325)\dpicstop
\dpicdraw (0.525,0.325)
 --(0.525,0)\dpicstop
\draw (0.608333,0.35) node[right=-1.5bp]{$ C_a$};
\dpicdraw[fill=black](0.525,0) circle (0.007874in)\dpicstop
\dpicdraw (0.525,0.7)
 --(0.75,0.7)
 --(0.770833,0.741667)
 --(0.8125,0.658333)
 --(0.854167,0.741667)
 --(0.895833,0.658333)
 --(0.9375,0.741667)
 --(0.979167,0.658333)
 --(1,0.7)
 --(1.225,0.7)\dpicstop
\draw (0.875,0.741667) node[above=-1.5bp]{$ R_{w}$};
\dpicdraw (1.225,0)
 --(1.225,0.225)\dpicstop
\dpicdraw (1.225,0.35) circle (0.049213in)\dpicstop
\draw (1.225,0.2875) node{$_-$};
\draw (1.225,0.4125) node{$_+$};
\dpicdraw (1.225,0.475)
 --(1.225,0.7)\dpicstop
\draw (1.35,0.35) node[right=-1.5bp]{$ T_r$};
\dpicdraw (1.225,0)
 --(0,0)\dpicstop
\end{tikzpicture}
\caption{Model~A TEC. The components $C_a$, $R_w$, $Q$ and $T_r$  respectively represent the lumped thermal mass of the freezer interior, thermal resistance of the envelope, the heat extracted from the freezer interior and the room at constant temperature. $V_a$ is the model output, \ie\ the freezer temperature.}\label{fig:modelA:circuit}
\end{figure}

\subsubsection{Model residuals analysis}
\citefig{fig:modelA:residuals} shows the logarithm of the absolute value of Model~A residuals autocorrelation function. The logarithm is to allow a clear separation between small values. Checking for autocorrelation in the model residuals is of importance to assess whether the model is able to capture all the dynamics contained in the training data set. This analysis consists in a visual comparison between the autocorrelation functions of model residuals and white noise (uncorrelated by definition). The latter is shown with the dashed line and should be intended as the threshold above which the model residuals are correlated in time. In the case of Model~A, residuals are significantly correlated, an evidence that indicates poor model performance.

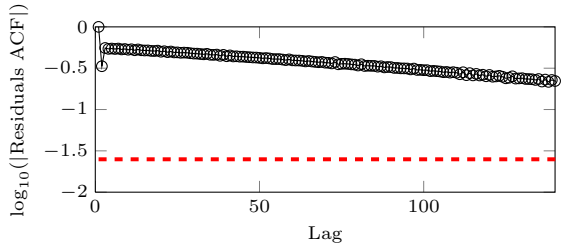
\begin{figure}[!ht]
\centering
\scriptsize
%
\begin{tikzpicture}

\begin{axis}[%
width=\matlabfigurewidth,
height=0.362327188940092\matlabfigurewidth,
scale only axis,
xmin=0,
xmax=140,
xlabel={Lag},
ymin=-2,
ymax=0,
ylabel={$\log_{10}(|\text{Residuals ACF}|)$}
]
\addplot [color=black,solid,mark=o,mark options={solid},forget plot]
  table[row sep=crcr]{%
1	-9.64327466553287e-17\\
2	-0.476162147611425\\
3	-0.255238381981706\\
4	-0.26601876511906\\
5	-0.26222561067303\\
6	-0.267472013143127\\
7	-0.262309558745207\\
8	-0.272811364632842\\
9	-0.267995165403784\\
10	-0.276551248947047\\
11	-0.271126329289872\\
12	-0.282821887170724\\
13	-0.273129039000015\\
14	-0.28315898892087\\
15	-0.285735173931182\\
16	-0.284714943339703\\
17	-0.290744394062486\\
18	-0.292118995880827\\
19	-0.287484562894492\\
20	-0.30084528128906\\
21	-0.294826593400558\\
22	-0.306143094800415\\
23	-0.298281705957013\\
24	-0.309136028956914\\
25	-0.308613682961881\\
26	-0.313630972187382\\
27	-0.311359614663318\\
28	-0.313361819549111\\
29	-0.319106090922386\\
30	-0.323325764129125\\
31	-0.323885895671044\\
32	-0.328520912949554\\
33	-0.332011126980482\\
34	-0.326169937917314\\
35	-0.337457650916968\\
36	-0.338844224816081\\
37	-0.335290335394505\\
38	-0.350090662604982\\
39	-0.338431944368497\\
40	-0.355353093152202\\
41	-0.345049706753293\\
42	-0.357833506096377\\
43	-0.355800892078654\\
44	-0.360664485068772\\
45	-0.360380226080702\\
46	-0.364522769296826\\
47	-0.367101132281942\\
48	-0.370216761568965\\
49	-0.371318634573251\\
50	-0.375181809995642\\
51	-0.380680479400358\\
52	-0.376945948911781\\
53	-0.381743129425198\\
54	-0.389375571719009\\
55	-0.389824448192642\\
56	-0.389701099131227\\
57	-0.395237219561038\\
58	-0.402166007943034\\
59	-0.393471821936268\\
60	-0.405109461137333\\
61	-0.406695506208093\\
62	-0.407168842013242\\
63	-0.410867547237018\\
64	-0.417769730059129\\
65	-0.412573553498266\\
66	-0.419255426293484\\
67	-0.423865179852414\\
68	-0.427848995962028\\
69	-0.423701907410967\\
70	-0.433080563261472\\
71	-0.434625686082908\\
72	-0.442252807780032\\
73	-0.425231548973637\\
74	-0.452777911911598\\
75	-0.446886653593782\\
76	-0.445758448554952\\
77	-0.448611009456502\\
78	-0.453914600242908\\
79	-0.456788799193931\\
80	-0.467563460091913\\
81	-0.452032440686929\\
82	-0.473286199562033\\
83	-0.470112451809264\\
84	-0.471711209747368\\
85	-0.472197694493026\\
86	-0.4816632347395\\
87	-0.47669918209971\\
88	-0.490244313672847\\
89	-0.485353736124759\\
90	-0.486847001065441\\
91	-0.496862806492928\\
92	-0.496706081076604\\
93	-0.501827866850831\\
94	-0.502387166971522\\
95	-0.503810023740608\\
96	-0.514612411648629\\
97	-0.504475520469229\\
98	-0.52127072713207\\
99	-0.519805162168722\\
100	-0.524221052113186\\
101	-0.523378297206119\\
102	-0.535594838236721\\
103	-0.532288331526435\\
104	-0.53610837914607\\
105	-0.540067139811853\\
106	-0.548455427696411\\
107	-0.545270528179976\\
108	-0.553619554851456\\
109	-0.547860226938026\\
110	-0.554713660877274\\
111	-0.573234578862409\\
112	-0.549024975699016\\
113	-0.584409254345882\\
114	-0.559502172478901\\
115	-0.587037185522179\\
116	-0.568896002328744\\
117	-0.58800411253939\\
118	-0.578993587851482\\
119	-0.597737835472944\\
120	-0.581554130045217\\
121	-0.603144376552816\\
122	-0.599445319931855\\
123	-0.592108483887204\\
124	-0.603415770515703\\
125	-0.622268096646101\\
126	-0.615738970760534\\
127	-0.598050573705461\\
128	-0.625682950029781\\
129	-0.623454014968826\\
130	-0.621154521412802\\
131	-0.630177476642386\\
132	-0.63127414326004\\
133	-0.633680344433868\\
134	-0.641961745373216\\
135	-0.631076769880443\\
136	-0.661408304601696\\
137	-0.637750594879995\\
138	-0.665629838627549\\
139	-0.642553210058948\\
140	-0.654826136201009\\
141	-0.664591269716846\\
142	-0.653989665774063\\
143	-0.676261721884042\\
144	-0.670259014253517\\
145	-0.663965572804108\\
146	-0.676228271493672\\
147	-0.67511018467894\\
148	-0.664178200693797\\
149	-0.706883668978797\\
150	-0.678517776931781\\
151	-0.681661563506975\\
152	-0.697196559117793\\
153	-0.695219607052666\\
154	-0.685002133415103\\
155	-0.706529616783872\\
156	-0.697783112241561\\
157	-0.709960262708369\\
158	-0.6923271679958\\
159	-0.726581404686603\\
160	-0.712898404125051\\
161	-0.704069096354843\\
162	-0.71688173903924\\
163	-0.728093230511488\\
164	-0.713254792669884\\
165	-0.729252176906782\\
166	-0.727450991944821\\
167	-0.730607853495952\\
168	-0.740239885248052\\
169	-0.728572915300671\\
170	-0.734909188762414\\
171	-0.753634674087514\\
172	-0.736856262865325\\
173	-0.747728165690357\\
174	-0.74661404261733\\
175	-0.774467797474038\\
176	-0.726592941810817\\
177	-0.774951411322046\\
178	-0.761387852646409\\
179	-0.760192034975223\\
180	-0.770053911756145\\
181	-0.763311082475969\\
182	-0.777016400153442\\
183	-0.776788377275374\\
184	-0.77915294054986\\
185	-0.773374616056068\\
186	-0.798760997790467\\
187	-0.779947608715555\\
188	-0.788738911962554\\
189	-0.788206167891118\\
190	-0.814325725892873\\
191	-0.79306279553991\\
192	-0.805019920787967\\
193	-0.790414427846412\\
};
\addplot [color=red,dashed,line width=1.5pt,forget plot]
  table[row sep=crcr]{%
1	-1.60205999132796\\
2	-1.60205999132796\\
3	-1.60205999132796\\
4	-1.60205999132796\\
5	-1.60205999132796\\
6	-1.60205999132796\\
7	-1.60205999132796\\
8	-1.60205999132796\\
9	-1.60205999132796\\
10	-1.60205999132796\\
11	-1.60205999132796\\
12	-1.60205999132796\\
13	-1.60205999132796\\
14	-1.60205999132796\\
15	-1.60205999132796\\
16	-1.60205999132796\\
17	-1.60205999132796\\
18	-1.60205999132796\\
19	-1.60205999132796\\
20	-1.60205999132796\\
21	-1.60205999132796\\
22	-1.60205999132796\\
23	-1.60205999132796\\
24	-1.60205999132796\\
25	-1.60205999132796\\
26	-1.60205999132796\\
27	-1.60205999132796\\
28	-1.60205999132796\\
29	-1.60205999132796\\
30	-1.60205999132796\\
31	-1.60205999132796\\
32	-1.60205999132796\\
33	-1.60205999132796\\
34	-1.60205999132796\\
35	-1.60205999132796\\
36	-1.60205999132796\\
37	-1.60205999132796\\
38	-1.60205999132796\\
39	-1.60205999132796\\
40	-1.60205999132796\\
41	-1.60205999132796\\
42	-1.60205999132796\\
43	-1.60205999132796\\
44	-1.60205999132796\\
45	-1.60205999132796\\
46	-1.60205999132796\\
47	-1.60205999132796\\
48	-1.60205999132796\\
49	-1.60205999132796\\
50	-1.60205999132796\\
51	-1.60205999132796\\
52	-1.60205999132796\\
53	-1.60205999132796\\
54	-1.60205999132796\\
55	-1.60205999132796\\
56	-1.60205999132796\\
57	-1.60205999132796\\
58	-1.60205999132796\\
59	-1.60205999132796\\
60	-1.60205999132796\\
61	-1.60205999132796\\
62	-1.60205999132796\\
63	-1.60205999132796\\
64	-1.60205999132796\\
65	-1.60205999132796\\
66	-1.60205999132796\\
67	-1.60205999132796\\
68	-1.60205999132796\\
69	-1.60205999132796\\
70	-1.60205999132796\\
71	-1.60205999132796\\
72	-1.60205999132796\\
73	-1.60205999132796\\
74	-1.60205999132796\\
75	-1.60205999132796\\
76	-1.60205999132796\\
77	-1.60205999132796\\
78	-1.60205999132796\\
79	-1.60205999132796\\
80	-1.60205999132796\\
81	-1.60205999132796\\
82	-1.60205999132796\\
83	-1.60205999132796\\
84	-1.60205999132796\\
85	-1.60205999132796\\
86	-1.60205999132796\\
87	-1.60205999132796\\
88	-1.60205999132796\\
89	-1.60205999132796\\
90	-1.60205999132796\\
91	-1.60205999132796\\
92	-1.60205999132796\\
93	-1.60205999132796\\
94	-1.60205999132796\\
95	-1.60205999132796\\
96	-1.60205999132796\\
97	-1.60205999132796\\
98	-1.60205999132796\\
99	-1.60205999132796\\
100	-1.60205999132796\\
101	-1.60205999132796\\
102	-1.60205999132796\\
103	-1.60205999132796\\
104	-1.60205999132796\\
105	-1.60205999132796\\
106	-1.60205999132796\\
107	-1.60205999132796\\
108	-1.60205999132796\\
109	-1.60205999132796\\
110	-1.60205999132796\\
111	-1.60205999132796\\
112	-1.60205999132796\\
113	-1.60205999132796\\
114	-1.60205999132796\\
115	-1.60205999132796\\
116	-1.60205999132796\\
117	-1.60205999132796\\
118	-1.60205999132796\\
119	-1.60205999132796\\
120	-1.60205999132796\\
121	-1.60205999132796\\
122	-1.60205999132796\\
123	-1.60205999132796\\
124	-1.60205999132796\\
125	-1.60205999132796\\
126	-1.60205999132796\\
127	-1.60205999132796\\
128	-1.60205999132796\\
129	-1.60205999132796\\
130	-1.60205999132796\\
131	-1.60205999132796\\
132	-1.60205999132796\\
133	-1.60205999132796\\
134	-1.60205999132796\\
135	-1.60205999132796\\
136	-1.60205999132796\\
137	-1.60205999132796\\
138	-1.60205999132796\\
139	-1.60205999132796\\
140	-1.60205999132796\\
141	-1.60205999132796\\
142	-1.60205999132796\\
143	-1.60205999132796\\
144	-1.60205999132796\\
145	-1.60205999132796\\
146	-1.60205999132796\\
147	-1.60205999132796\\
148	-1.60205999132796\\
149	-1.60205999132796\\
150	-1.60205999132796\\
151	-1.60205999132796\\
152	-1.60205999132796\\
153	-1.60205999132796\\
154	-1.60205999132796\\
155	-1.60205999132796\\
156	-1.60205999132796\\
157	-1.60205999132796\\
158	-1.60205999132796\\
159	-1.60205999132796\\
160	-1.60205999132796\\
161	-1.60205999132796\\
162	-1.60205999132796\\
163	-1.60205999132796\\
164	-1.60205999132796\\
165	-1.60205999132796\\
166	-1.60205999132796\\
167	-1.60205999132796\\
168	-1.60205999132796\\
169	-1.60205999132796\\
170	-1.60205999132796\\
171	-1.60205999132796\\
172	-1.60205999132796\\
173	-1.60205999132796\\
174	-1.60205999132796\\
175	-1.60205999132796\\
176	-1.60205999132796\\
177	-1.60205999132796\\
178	-1.60205999132796\\
179	-1.60205999132796\\
180	-1.60205999132796\\
181	-1.60205999132796\\
182	-1.60205999132796\\
183	-1.60205999132796\\
184	-1.60205999132796\\
185	-1.60205999132796\\
186	-1.60205999132796\\
187	-1.60205999132796\\
188	-1.60205999132796\\
189	-1.60205999132796\\
190	-1.60205999132796\\
191	-1.60205999132796\\
192	-1.60205999132796\\
193	-1.60205999132796\\
};
\end{axis}
\end{tikzpicture}%
\caption{The autocorrelation of Model~A residuals (full line) and white noise (dashed line).}
\label{fig:modelA:residuals}
\end{figure}

\subsection{Model~B (linear second order)}\label{sec:modelB}
\subsubsection{Model formulation}
From the measurements, it was possible to observe a time delay between the activation of the freezer compressor and initial temperature decay. This is because the thermal inertia of the freezer cold heat exchanger, that needs to cool down before being able to extract heat from the freezer interior. To account for such an effect, Model~B has an additional RC branch with respect to the previous model. From \citefig{fig:modelB:circuit}, the new components $C_e$ and $R_e$ respectively represent the thermal mass of the heat exchanger and the thermal contact resistance between the exchanger and the rest of the freezer cold mass. 
The state and input vectors of Model~B are as
\renewcommand*{\arraystretch}{1.3}
\begin{align}
\begin{aligned}
\boldsymbol{x}^T &= \begin{bmatrix} V_a & V_e \end{bmatrix} \\
\boldsymbol{u}^T &= \begin{bmatrix} T_r & P \end{bmatrix},
\end{aligned}
\end{align}
while the stochastic state space matrices are:
\begin{align}
A &= 
    \begin{bmatrix}
      -\frac{1}{C_a R_w} - \frac{1}{C_a R_e} &  \frac{1}{C_a R_e} \\
      \frac{1}{C_e R_e} & - \frac{1}{C_e R_e}
    \end{bmatrix} \\
B &= 
    \begin{bmatrix}
    \frac{1}{C_a R_w} & 0 \\
    0 & -\frac{\text{COP}}{C_e} 
    \end{bmatrix}\\
W &= \text{diag}(w_0, w_1)\\
C &= \begin{bmatrix} 1 & 0 \end{bmatrix}
\end{align}
where $\text{diag}(\cdot)$ denotes the diagonal matrix with elements as in the argument list. The values of the estimated parameters are shown in \citetable{tab:modelB:pars}.

\begin{figure}[ht]
\centering
\small
\begin{tikzpicture}[scale=2.54]
\ifx\dpiclw\undefined\newdimen\dpiclw\fi
\global\def\dpicdraw{\draw[line width=\dpiclw]}
\global\def\dpicstop{;}
\dpiclw=0.8bp
\dpiclw=0.8bp
\dpiclw=0.85bp
\dpicdraw (0,0.7)
 --(0,0.475)\dpicstop
\dpicdraw (0,0.35) circle (0.049213in)\dpicstop
\dpicdraw (0,0.44375)
 --(0,0.3625)\dpicstop
\filldraw[line width=0bp](-0.026562,0.3625)
 --(0,0.25625)
 --(0.026563,0.3625) --cycle
\dpicstop
\dpicdraw (0,0.225)
 --(0,0)\dpicstop
\draw (0.125,0.35) node[right=-1.5bp]{$ Q$};
\dpicdraw (0,0.7)
 --(0.525,0.7)\dpicstop
\draw (0.525,0.85) node{$V_e$};
\dpicdraw[fill=black](0.525,0.7) circle (0.007874in)\dpicstop
\dpicdraw (0.525,0.7)
 --(0.525,0.375)\dpicstop
\dpicdraw (0.441667,0.375)
 --(0.608333,0.375)\dpicstop
\dpicdraw (0.441667,0.325)
 --(0.608333,0.325)\dpicstop
\dpicdraw (0.525,0.325)
 --(0.525,0)\dpicstop
\draw (0.608333,0.35) node[right=-1.5bp]{$ C_e$};
\dpicdraw[fill=black](0.525,0) circle (0.007874in)\dpicstop
\dpicdraw (0.525,0.7)
 --(0.75,0.7)
 --(0.770833,0.741667)
 --(0.8125,0.658333)
 --(0.854167,0.741667)
 --(0.895833,0.658333)
 --(0.9375,0.741667)
 --(0.979167,0.658333)
 --(1,0.7)
 --(1.225,0.7)\dpicstop
\draw (0.875,0.741667) node[above=-1.5bp]{$ R_{e}$};
\draw (1.225,0.85) node{$V_a$};
\dpicdraw[fill=black](1.225,0.7) circle (0.007874in)\dpicstop
\dpicdraw (1.225,0.7)
 --(1.225,0.375)\dpicstop
\dpicdraw (1.141667,0.375)
 --(1.308333,0.375)\dpicstop
\dpicdraw (1.141667,0.325)
 --(1.308333,0.325)\dpicstop
\dpicdraw (1.225,0.325)
 --(1.225,0)\dpicstop
\draw (1.308333,0.35) node[right=-1.5bp]{$ C_a$};
\dpicdraw[fill=black](1.225,0) circle (0.007874in)\dpicstop
\dpicdraw (1.225,0.7)
 --(1.45,0.7)
 --(1.470833,0.741667)
 --(1.5125,0.658333)
 --(1.554167,0.741667)
 --(1.595833,0.658333)
 --(1.6375,0.741667)
 --(1.679167,0.658333)
 --(1.7,0.7)
 --(1.925,0.7)\dpicstop
\draw (1.575,0.741667) node[above=-1.5bp]{$ R_{w}$};
\dpicdraw (1.925,0)
 --(1.925,0.225)\dpicstop
\dpicdraw (1.925,0.35) circle (0.049213in)\dpicstop
\draw (1.925,0.2875) node{$_-$};
\draw (1.925,0.4125) node{$_+$};
\dpicdraw (1.925,0.475)
 --(1.925,0.7)\dpicstop
\draw (2.05,0.35) node[right=-1.5bp]{$ T_r$};
\dpicdraw (1.925,0)
 --(0,0)\dpicstop
\end{tikzpicture}
\caption{Model~B TEC. The additional RC branch decouples between the thermal masses of the freezer interior and heat exchanger.}\label{fig:modelB:circuit}
\end{figure}
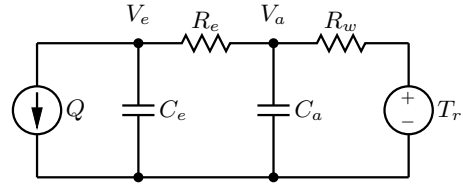

\subsubsection{Model identification and validation}
As visible from \citefig{fig:modelB:residuals}, Model~B residuals are significantly less correlated than in the previous case. This indicates that the additional state absorbed a part of the dynamics that were left unexplained by the previous model. 
We now apply the procedure described in \ref{methods:model_expansion} to validate the model extension. The deviance \eqref{eq:deviance} calculated using the log-likelihood values in Table \ref{tab:modelA:pars} and \ref{tab:modelB:pars} is $D=10665$. The p-value \eqref{eq:pvalue} of the current model extension is very close to zero, indeed below the 5\% threshold. The null hypothesis is therefore rejected and Model~B is considered a valid model.

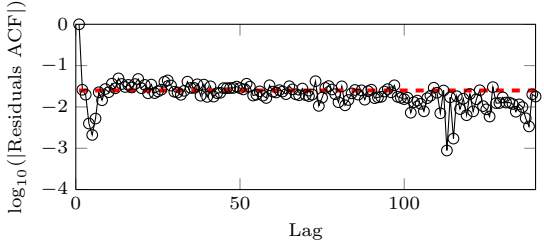
\begin{figure}[!ht]
\centering
\scriptsize
%
\begin{tikzpicture}

\begin{axis}[%
width=\matlabfigurewidth,
height=0.362327188940092\matlabfigurewidth,
scale only axis,
xmin=0,
xmax=140,
xlabel={Lag},
ymin=-4,
ymax=0,
ylabel={$\log_{10}(|\text{Residuals ACF}|)$}
]
\addplot [color=black,solid,mark=o,mark options={solid},forget plot]
  table[row sep=crcr]{%
1	0\\
2	-1.58442861064302\\
3	-1.70362971445443\\
4	-2.40143596237146\\
5	-2.67750051536018\\
6	-2.28147434169021\\
7	-1.64506804518142\\
8	-1.83776599581628\\
9	-1.56658066016937\\
10	-1.64260888132045\\
11	-1.43713193363207\\
12	-1.54923669022289\\
13	-1.30929222059779\\
14	-1.44492891453451\\
15	-1.51652558709783\\
16	-1.46667244137937\\
17	-1.52901163411797\\
18	-1.44659741848532\\
19	-1.32348762633448\\
20	-1.54706596892397\\
21	-1.46615035780278\\
22	-1.66859187541188\\
23	-1.45923293190065\\
24	-1.66508826940856\\
25	-1.62317086350613\\
26	-1.58383379422307\\
27	-1.39128417365897\\
28	-1.35827984017738\\
29	-1.48212492749113\\
30	-1.63163447145673\\
31	-1.64888929411719\\
32	-1.70648487907885\\
33	-1.61021445953081\\
34	-1.38798475316772\\
35	-1.54068315398658\\
36	-1.53932796808372\\
37	-1.45241048987448\\
38	-1.727921757855\\
39	-1.45965762173844\\
40	-1.75647426980182\\
41	-1.50069310465683\\
42	-1.74939635925518\\
43	-1.66541186671519\\
44	-1.65719163338485\\
45	-1.54444167678402\\
46	-1.54565746852106\\
47	-1.54197489200815\\
48	-1.53948342024353\\
49	-1.50812237835011\\
50	-1.54427641170692\\
51	-1.57605940526105\\
52	-1.43827006617523\\
53	-1.50840012223467\\
54	-1.72379137571348\\
55	-1.70251247001696\\
56	-1.62275636970146\\
57	-1.72521716254893\\
58	-1.78326779551635\\
59	-1.47496074494899\\
60	-1.63025606211831\\
61	-1.65043718054984\\
62	-1.59341045684437\\
63	-1.6263030929456\\
64	-1.69039528834935\\
65	-1.49805580998782\\
66	-1.57414459703527\\
67	-1.69423641434991\\
68	-1.71588575197584\\
69	-1.55505637758035\\
70	-1.71341761402788\\
71	-1.73809755214169\\
72	-1.70080828639781\\
73	-1.37709141177128\\
74	-1.97473955462771\\
75	-1.77443137937776\\
76	-1.53996996750422\\
77	-1.4951014839701\\
78	-1.57788721323295\\
79	-1.69096871961981\\
80	-1.88648765861822\\
81	-1.50173765838083\\
82	-1.9606853603154\\
83	-1.83168794996428\\
84	-1.67169699406543\\
85	-1.58929129457987\\
86	-1.70215523122156\\
87	-1.59848513799004\\
88	-1.82423183711335\\
89	-1.61915925304802\\
90	-1.58284147381304\\
91	-1.78736809958994\\
92	-1.76734445003345\\
93	-1.75572215162634\\
94	-1.62733653688364\\
95	-1.56731180437027\\
96	-1.64676474878038\\
97	-1.47601778845519\\
98	-1.75549718953295\\
99	-1.78004372427359\\
100	-1.81084513670501\\
101	-1.77676292585145\\
102	-2.13999654539477\\
103	-1.88465128803713\\
104	-1.84664546072126\\
105	-1.92381949713701\\
106	-2.1056230933347\\
107	-1.78809993198271\\
108	-1.72932724071104\\
109	-1.53567967223759\\
110	-1.65674136268571\\
111	-2.15305598374608\\
112	-1.60313338861084\\
113	-3.0542047041622\\
114	-1.76711316426284\\
115	-2.76641718047904\\
116	-1.74068359072899\\
117	-2.05114685236589\\
118	-1.80079558664837\\
119	-2.20017265876209\\
120	-1.68778651859706\\
121	-2.10813636807228\\
122	-1.81284271586707\\
123	-1.59617849331778\\
124	-1.98015223100332\\
125	-2.05775746325078\\
126	-2.23270818461923\\
127	-1.52002643468575\\
128	-1.90764362991174\\
129	-1.90494284119451\\
130	-1.76774784146518\\
131	-1.89364202254238\\
132	-1.89418546370968\\
133	-1.927861973372\\
134	-2.1153499993584\\
135	-1.93804425691816\\
136	-2.00411302132344\\
137	-2.27441967089891\\
138	-2.47164587278349\\
139	-1.69457863714191\\
140	-1.74819261128769\\
141	-1.96132410342112\\
142	-1.83051567162595\\
143	-3.52238931643455\\
144	-2.09045863926342\\
145	-1.67744069365327\\
146	-1.72377691032118\\
147	-1.64421736433749\\
148	-1.62624220866432\\
149	-2.08271967584803\\
150	-1.99314302607293\\
151	-1.82873930464466\\
152	-2.18164749040706\\
153	-1.91583795013728\\
154	-1.65972607732536\\
155	-2.03748070872896\\
156	-1.90361314549918\\
157	-2.05980366712351\\
158	-1.77361838327995\\
159	-2.38674515237612\\
160	-2.08755535732512\\
161	-1.65475783042235\\
162	-1.80484873996353\\
163	-2.04655739665165\\
164	-1.74120139568383\\
165	-2.01694468855142\\
166	-2.04778609997026\\
167	-2.12966281477601\\
168	-2.24622387296508\\
169	-1.77738499807547\\
170	-1.86849654390592\\
171	-2.53620874408445\\
172	-1.86478443633858\\
173	-1.98432458829736\\
174	-2.10095576819757\\
175	-3.0579819854723\\
176	-1.55435760414725\\
177	-2.43183616404819\\
178	-2.08261968027702\\
179	-1.8805933107848\\
180	-1.94345569332487\\
181	-1.82053691317948\\
182	-2.10121081621241\\
183	-2.08273843710715\\
184	-1.98488242980804\\
185	-1.88026625558402\\
186	-2.69671894734479\\
187	-1.85267804697404\\
188	-1.87320817748838\\
189	-2.01533055363225\\
190	-2.51742902415488\\
191	-1.92009857557766\\
192	-1.79160918328176\\
193	-1.62319264174465\\
};
\addplot [color=red,dashed,line width=1.5pt,forget plot]
  table[row sep=crcr]{%
1	-1.60205999132796\\
2	-1.60205999132796\\
3	-1.60205999132796\\
4	-1.60205999132796\\
5	-1.60205999132796\\
6	-1.60205999132796\\
7	-1.60205999132796\\
8	-1.60205999132796\\
9	-1.60205999132796\\
10	-1.60205999132796\\
11	-1.60205999132796\\
12	-1.60205999132796\\
13	-1.60205999132796\\
14	-1.60205999132796\\
15	-1.60205999132796\\
16	-1.60205999132796\\
17	-1.60205999132796\\
18	-1.60205999132796\\
19	-1.60205999132796\\
20	-1.60205999132796\\
21	-1.60205999132796\\
22	-1.60205999132796\\
23	-1.60205999132796\\
24	-1.60205999132796\\
25	-1.60205999132796\\
26	-1.60205999132796\\
27	-1.60205999132796\\
28	-1.60205999132796\\
29	-1.60205999132796\\
30	-1.60205999132796\\
31	-1.60205999132796\\
32	-1.60205999132796\\
33	-1.60205999132796\\
34	-1.60205999132796\\
35	-1.60205999132796\\
36	-1.60205999132796\\
37	-1.60205999132796\\
38	-1.60205999132796\\
39	-1.60205999132796\\
40	-1.60205999132796\\
41	-1.60205999132796\\
42	-1.60205999132796\\
43	-1.60205999132796\\
44	-1.60205999132796\\
45	-1.60205999132796\\
46	-1.60205999132796\\
47	-1.60205999132796\\
48	-1.60205999132796\\
49	-1.60205999132796\\
50	-1.60205999132796\\
51	-1.60205999132796\\
52	-1.60205999132796\\
53	-1.60205999132796\\
54	-1.60205999132796\\
55	-1.60205999132796\\
56	-1.60205999132796\\
57	-1.60205999132796\\
58	-1.60205999132796\\
59	-1.60205999132796\\
60	-1.60205999132796\\
61	-1.60205999132796\\
62	-1.60205999132796\\
63	-1.60205999132796\\
64	-1.60205999132796\\
65	-1.60205999132796\\
66	-1.60205999132796\\
67	-1.60205999132796\\
68	-1.60205999132796\\
69	-1.60205999132796\\
70	-1.60205999132796\\
71	-1.60205999132796\\
72	-1.60205999132796\\
73	-1.60205999132796\\
74	-1.60205999132796\\
75	-1.60205999132796\\
76	-1.60205999132796\\
77	-1.60205999132796\\
78	-1.60205999132796\\
79	-1.60205999132796\\
80	-1.60205999132796\\
81	-1.60205999132796\\
82	-1.60205999132796\\
83	-1.60205999132796\\
84	-1.60205999132796\\
85	-1.60205999132796\\
86	-1.60205999132796\\
87	-1.60205999132796\\
88	-1.60205999132796\\
89	-1.60205999132796\\
90	-1.60205999132796\\
91	-1.60205999132796\\
92	-1.60205999132796\\
93	-1.60205999132796\\
94	-1.60205999132796\\
95	-1.60205999132796\\
96	-1.60205999132796\\
97	-1.60205999132796\\
98	-1.60205999132796\\
99	-1.60205999132796\\
100	-1.60205999132796\\
101	-1.60205999132796\\
102	-1.60205999132796\\
103	-1.60205999132796\\
104	-1.60205999132796\\
105	-1.60205999132796\\
106	-1.60205999132796\\
107	-1.60205999132796\\
108	-1.60205999132796\\
109	-1.60205999132796\\
110	-1.60205999132796\\
111	-1.60205999132796\\
112	-1.60205999132796\\
113	-1.60205999132796\\
114	-1.60205999132796\\
115	-1.60205999132796\\
116	-1.60205999132796\\
117	-1.60205999132796\\
118	-1.60205999132796\\
119	-1.60205999132796\\
120	-1.60205999132796\\
121	-1.60205999132796\\
122	-1.60205999132796\\
123	-1.60205999132796\\
124	-1.60205999132796\\
125	-1.60205999132796\\
126	-1.60205999132796\\
127	-1.60205999132796\\
128	-1.60205999132796\\
129	-1.60205999132796\\
130	-1.60205999132796\\
131	-1.60205999132796\\
132	-1.60205999132796\\
133	-1.60205999132796\\
134	-1.60205999132796\\
135	-1.60205999132796\\
136	-1.60205999132796\\
137	-1.60205999132796\\
138	-1.60205999132796\\
139	-1.60205999132796\\
140	-1.60205999132796\\
141	-1.60205999132796\\
142	-1.60205999132796\\
143	-1.60205999132796\\
144	-1.60205999132796\\
145	-1.60205999132796\\
146	-1.60205999132796\\
147	-1.60205999132796\\
148	-1.60205999132796\\
149	-1.60205999132796\\
150	-1.60205999132796\\
151	-1.60205999132796\\
152	-1.60205999132796\\
153	-1.60205999132796\\
154	-1.60205999132796\\
155	-1.60205999132796\\
156	-1.60205999132796\\
157	-1.60205999132796\\
158	-1.60205999132796\\
159	-1.60205999132796\\
160	-1.60205999132796\\
161	-1.60205999132796\\
162	-1.60205999132796\\
163	-1.60205999132796\\
164	-1.60205999132796\\
165	-1.60205999132796\\
166	-1.60205999132796\\
167	-1.60205999132796\\
168	-1.60205999132796\\
169	-1.60205999132796\\
170	-1.60205999132796\\
171	-1.60205999132796\\
172	-1.60205999132796\\
173	-1.60205999132796\\
174	-1.60205999132796\\
175	-1.60205999132796\\
176	-1.60205999132796\\
177	-1.60205999132796\\
178	-1.60205999132796\\
179	-1.60205999132796\\
180	-1.60205999132796\\
181	-1.60205999132796\\
182	-1.60205999132796\\
183	-1.60205999132796\\
184	-1.60205999132796\\
185	-1.60205999132796\\
186	-1.60205999132796\\
187	-1.60205999132796\\
188	-1.60205999132796\\
189	-1.60205999132796\\
190	-1.60205999132796\\
191	-1.60205999132796\\
192	-1.60205999132796\\
193	-1.60205999132796\\
};
\end{axis}
\end{tikzpicture}%
\caption{The autocorrelation of Model~B residuals (full line) and white noise (dashed line).}
\label{fig:modelB:residuals}
\end{figure}

\subsection{Model~C (linear third order)}\label{sec:modelC}
\subsubsection{Model formulation}
As shown in \citefig{fig:model:circuitC} and with respect to the previous model, an additional state $C_w$ is added to decouple between the thermal masses of the freezer envelope and air content. The resistor $R_a$ models the thermal resistance between the two.
Model~C state and input vectors are
\begin{align}
\begin{aligned}
\boldsymbol{x}^T &= \begin{bmatrix} V_e & V_a & V_w \end{bmatrix} \\
\boldsymbol{u}^T &= \begin{bmatrix} T_r & P \end{bmatrix},
\end{aligned}
\end{align}
while the stochastic state space matrices are as:
\begingroup
\setlength{\arraycolsep}{3pt}
\begin{align}
A &= \begin{bmatrix}
      \frac{-1}{C_e R_e} &  \frac{1}{C_e R_e} &  0 \\
      \frac{1}{C_a R_e}  & \frac{-1}{C_a R_a}  + \frac{-1}{C_a R_e} &  \frac{1}{C_a R_a} \\
      0 & \frac{1}{C_w R_a} &  \frac{-1}{C_w R_w}  +  \frac{-1}{C_w R_a}
     \end{bmatrix}\\
B &= \begin{bmatrix}
      0 & -\frac{\text{COP}}{C_e} \\
      0 & 0 \\
      \frac{1}{C_w R_w} & 0
     \end{bmatrix} \\
W &= \text{diag}(w_0, w_1, w_2)\\
C &= \begin{bmatrix} 0 & 1 & 0 \end{bmatrix}
\end{align}
\endgroup

\begin{figure}[ht]
\centering
\small
\begin{tikzpicture}[scale=2.54]
\ifx\dpiclw\undefined\newdimen\dpiclw\fi
\global\def\dpicdraw{\draw[line width=\dpiclw]}
\global\def\dpicstop{;}
\dpiclw=0.8bp
\dpiclw=0.8bp
\dpiclw=0.85bp
\dpicdraw (0,0.7)
 --(0,0.475)\dpicstop
\dpicdraw (0,0.35) circle (0.049213in)\dpicstop
\dpicdraw (0,0.44375)
 --(0,0.3625)\dpicstop
\filldraw[line width=0bp](-0.026562,0.3625)
 --(0,0.25625)
 --(0.026563,0.3625) --cycle
\dpicstop
\dpicdraw (0,0.225)
 --(0,0)\dpicstop
\draw (0.125,0.35) node[right=-1.5bp]{$ Q$};
\dpicdraw (0,0.7)
 --(0.525,0.7)\dpicstop
\draw (0.525,0.85) node{$V_e$};
\dpicdraw[fill=black](0.525,0.7) circle (0.007874in)\dpicstop
\dpicdraw (0.525,0.7)
 --(0.525,0.375)\dpicstop
\dpicdraw (0.441667,0.375)
 --(0.608333,0.375)\dpicstop
\dpicdraw (0.441667,0.325)
 --(0.608333,0.325)\dpicstop
\dpicdraw (0.525,0.325)
 --(0.525,0)\dpicstop
\draw (0.608333,0.35) node[right=-1.5bp]{$ C_e$};
\dpicdraw[fill=black](0.525,0) circle (0.007874in)\dpicstop
\dpicdraw (0.525,0.7)
 --(0.6625,0.7)
 --(0.683333,0.741667)
 --(0.725,0.658333)
 --(0.766667,0.741667)
 --(0.808333,0.658333)
 --(0.85,0.741667)
 --(0.891667,0.658333)
 --(0.9125,0.7)
 --(1.05,0.7)\dpicstop
\draw (0.7875,0.741667) node[above=-1.5bp]{$ R_{e}$};
\draw (1.05,0.85) node{$V_a$};
\dpicdraw[fill=black](1.05,0.7) circle (0.007874in)\dpicstop
\dpicdraw (1.05,0.7)
 --(1.05,0.375)\dpicstop
\dpicdraw (0.966667,0.375)
 --(1.133333,0.375)\dpicstop
\dpicdraw (0.966667,0.325)
 --(1.133333,0.325)\dpicstop
\dpicdraw (1.05,0.325)
 --(1.05,0)\dpicstop
\draw (1.133333,0.35) node[right=-1.5bp]{$ C_a$};
\dpicdraw[fill=black](1.05,0) circle (0.007874in)\dpicstop
\dpicdraw (1.05,0.7)
 --(1.1875,0.7)
 --(1.208333,0.741667)
 --(1.25,0.658333)
 --(1.291667,0.741667)
 --(1.333333,0.658333)
 --(1.375,0.741667)
 --(1.416667,0.658333)
 --(1.4375,0.7)
 --(1.575,0.7)\dpicstop
\draw (1.3125,0.741667) node[above=-1.5bp]{$ R_{a}$};
\draw (1.575,0.85) node{$V_w$};
\dpicdraw[fill=black](1.575,0.7) circle (0.007874in)\dpicstop
\dpicdraw (1.575,0.7)
 --(1.575,0.375)\dpicstop
\dpicdraw (1.491667,0.375)
 --(1.658333,0.375)\dpicstop
\dpicdraw (1.491667,0.325)
 --(1.658333,0.325)\dpicstop
\dpicdraw (1.575,0.325)
 --(1.575,0)\dpicstop
\draw (1.658333,0.35) node[right=-1.5bp]{$ C_w$};
\dpicdraw[fill=black](1.575,0) circle (0.007874in)\dpicstop
\dpicdraw (1.575,0.7)
 --(1.7125,0.7)
 --(1.733333,0.741667)
 --(1.775,0.658333)
 --(1.816667,0.741667)
 --(1.858333,0.658333)
 --(1.9,0.741667)
 --(1.941667,0.658333)
 --(1.9625,0.7)
 --(2.1,0.7)\dpicstop
\draw (1.8375,0.741667) node[above=-1.5bp]{$ R_{w}$};
\dpicdraw (2.1,0)
 --(2.1,0.225)\dpicstop
\dpicdraw (2.1,0.35) circle (0.049213in)\dpicstop
\draw (2.1,0.2875) node{$_-$};
\draw (2.1,0.4125) node{$_+$};
\dpicdraw (2.1,0.475)
 --(2.1,0.7)\dpicstop
\draw (2.225,0.35) node[right=-1.5bp]{$ T_r$};
\dpicdraw (2.1,0)
 --(0,0)\dpicstop
\end{tikzpicture}
\caption{Model~C TEC. The freezer interior thermal mass is further decoupled by adding another RC branch.}\label{fig:model:circuitC}
\end{figure}
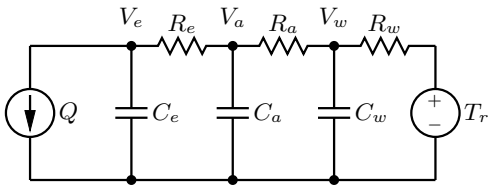

\subsubsection{Model identification and validation} \label{sec:modelCvalidation}
The autocorrelation of Model~C residuals in \citefig{fig:modelC:residuals} further improved with respect to the previous models, and only a few points are above the autocorrelation threshold. 
The deviance of the current model extension is $D=7.0$. The associated p-value is 3.0\%, that is below the 5\% significance level. The null hypothesis is rejected and Model~C is considered a valid model.

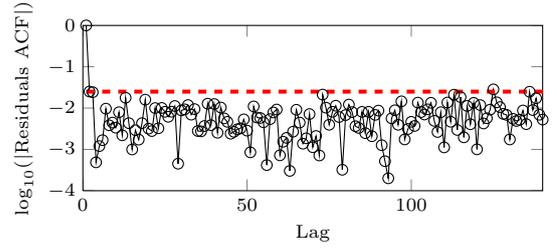
\begin{figure}[!ht]
\centering
\scriptsize
%
\begin{tikzpicture}

\begin{axis}[%
width=\matlabfigurewidth,
height=0.362327188940092\matlabfigurewidth,
scale only axis,
xmin=0,
xmax=140,
xlabel={Lag},
ymin=-4,
ymax=0,
ylabel={$\log_{10}(|\text{Residuals ACF}|)$}
]
\addplot [color=black,solid,mark=o,mark options={solid},forget plot]
  table[row sep=crcr]{%
1	0\\
2	-1.59964130696613\\
3	-1.61619166230759\\
4	-3.31329585979336\\
5	-2.91982893016816\\
6	-2.77634027325918\\
7	-2.01367762904864\\
8	-2.41668360897806\\
9	-2.35110926648633\\
10	-2.4742181708068\\
11	-2.1004209995578\\
12	-2.65936488269045\\
13	-1.75012757376144\\
14	-2.36520869164908\\
15	-3.00138097820412\\
16	-2.53577628907193\\
17	-2.76286731875556\\
18	-2.36495550029577\\
19	-1.79638001909039\\
20	-2.49756043181907\\
21	-2.55558332531234\\
22	-1.99965503501887\\
23	-2.48910278212143\\
24	-1.9969775316001\\
25	-2.09017488420791\\
26	-2.21089995256289\\
27	-2.08795420104353\\
28	-1.94987994415388\\
29	-3.34548008179017\\
30	-2.05616044884797\\
31	-2.03111052441994\\
32	-1.92680399440168\\
33	-2.15897269718755\\
34	-2.02455709810755\\
35	-2.55451422524641\\
36	-2.56320279299149\\
37	-2.43418586580955\\
38	-1.89376777853706\\
39	-2.41271534934111\\
40	-1.90222749124938\\
41	-2.59643649236262\\
42	-1.98629912060432\\
43	-2.24898104766609\\
44	-2.33314492466097\\
45	-2.6196424465425\\
46	-2.57435802383381\\
47	-2.51466377122388\\
48	-2.49147212588148\\
49	-2.27295207674255\\
50	-2.54576096460005\\
51	-3.07056737507086\\
52	-1.95985939347827\\
53	-2.22780599349903\\
54	-2.23927414973333\\
55	-2.34348783364567\\
56	-3.37910568973874\\
57	-2.26775859374107\\
58	-2.10593806181696\\
59	-2.03502953974871\\
60	-3.14319664930172\\
61	-2.81459754726051\\
62	-2.72068544960944\\
63	-3.52004851214717\\
64	-2.57125197538534\\
65	-2.04135961671128\\
66	-2.35018659946841\\
67	-2.8613136888125\\
68	-2.73125776401776\\
69	-2.14477615603767\\
70	-2.94697318011609\\
71	-2.67545866551008\\
72	-3.14692199682243\\
73	-1.67350345090017\\
74	-1.98941865151589\\
75	-2.40344283511459\\
76	-2.09054723331959\\
77	-1.94252272178677\\
78	-2.21654787482884\\
79	-3.49275435596309\\
80	-2.16904083041229\\
81	-1.91512437209431\\
82	-2.097895944694\\
83	-2.43855264241235\\
84	-2.48073054878086\\
85	-2.09192050463065\\
86	-2.60635762350478\\
87	-2.09101033810697\\
88	-2.68118969699884\\
89	-2.16459362343655\\
90	-2.06099673595969\\
91	-2.89818173351464\\
92	-3.28559877667838\\
93	-3.69962848453833\\
94	-2.24912008220754\\
95	-2.0534164994201\\
96	-2.40370974411595\\
97	-1.836869626658\\
98	-2.75805523278332\\
99	-2.50524016711844\\
100	-2.33262571903241\\
101	-2.43254627970737\\
102	-1.86795552792777\\
103	-2.0965095049507\\
104	-2.1563500512438\\
105	-2.03095707918532\\
106	-1.87241023467964\\
107	-2.30853505639191\\
108	-2.57824237782521\\
109	-2.09965911752703\\
110	-2.95101900141345\\
111	-1.8653747507452\\
112	-2.33025002073637\\
113	-1.67603559756337\\
114	-2.52877902458799\\
115	-1.73398651913786\\
116	-2.70810902936287\\
117	-1.94826204915983\\
118	-2.3917310585776\\
119	-1.87340544778042\\
120	-2.99805352015846\\
121	-1.92840323751848\\
122	-2.37215318479008\\
123	-2.24577572585096\\
124	-2.03816460876865\\
125	-1.54730079595418\\
126	-1.86018560316226\\
127	-1.97630688199349\\
128	-2.13983441468953\\
129	-2.1642150413553\\
130	-2.75938517394785\\
131	-2.26069386871301\\
132	-2.31188911588642\\
133	-2.28267137302829\\
134	-2.05855318253016\\
135	-2.38861174654031\\
136	-1.60421551772972\\
137	-2.04752729229774\\
138	-1.76303096743177\\
139	-2.16174381300926\\
140	-2.27789676169066\\
141	-3.1476329060629\\
142	-2.39150157505318\\
143	-1.98371530989963\\
144	-2.84138609380583\\
145	-1.92477486155492\\
146	-1.98828803903748\\
147	-1.83200875205249\\
148	-1.78370544540547\\
149	-1.82587848052134\\
150	-2.42146560298144\\
151	-2.07889745999653\\
152	-3.80407024559784\\
153	-2.29818724498807\\
154	-1.84191739287976\\
155	-2.85391804230298\\
156	-2.32612616613962\\
157	-3.00130956131712\\
158	-2.03417585868858\\
159	-1.92716530301573\\
160	-3.62536136803144\\
161	-1.85912856796186\\
162	-2.16483793246171\\
163	-3.32862896461143\\
164	-2.09033546680011\\
165	-3.03877344614111\\
166	-2.73054470005042\\
167	-2.45109261440738\\
168	-2.27787585744426\\
169	-2.22716384833729\\
170	-2.51477034810386\\
171	-2.13586521609893\\
172	-2.42367535549633\\
173	-3.17493717320138\\
174	-2.77646920723109\\
175	-1.97599488257563\\
176	-1.73875488974966\\
177	-2.21248598209955\\
178	-2.79649990686712\\
179	-2.50021168425077\\
180	-2.90641720753515\\
181	-2.30094303159158\\
182	-2.66449441127591\\
183	-2.74386301862353\\
184	-3.46643576062833\\
185	-2.49367735084344\\
186	-2.09734095098638\\
187	-2.3927945327915\\
188	-2.46106337417311\\
189	-4.14317677515601\\
190	-1.89755192412499\\
191	-2.60555631557551\\
192	-2.18050101541094\\
193	-1.84259111674562\\
};
\addplot [color=red,dashed,line width=1.5pt,forget plot]
  table[row sep=crcr]{%
1	-1.60205999132796\\
2	-1.60205999132796\\
3	-1.60205999132796\\
4	-1.60205999132796\\
5	-1.60205999132796\\
6	-1.60205999132796\\
7	-1.60205999132796\\
8	-1.60205999132796\\
9	-1.60205999132796\\
10	-1.60205999132796\\
11	-1.60205999132796\\
12	-1.60205999132796\\
13	-1.60205999132796\\
14	-1.60205999132796\\
15	-1.60205999132796\\
16	-1.60205999132796\\
17	-1.60205999132796\\
18	-1.60205999132796\\
19	-1.60205999132796\\
20	-1.60205999132796\\
21	-1.60205999132796\\
22	-1.60205999132796\\
23	-1.60205999132796\\
24	-1.60205999132796\\
25	-1.60205999132796\\
26	-1.60205999132796\\
27	-1.60205999132796\\
28	-1.60205999132796\\
29	-1.60205999132796\\
30	-1.60205999132796\\
31	-1.60205999132796\\
32	-1.60205999132796\\
33	-1.60205999132796\\
34	-1.60205999132796\\
35	-1.60205999132796\\
36	-1.60205999132796\\
37	-1.60205999132796\\
38	-1.60205999132796\\
39	-1.60205999132796\\
40	-1.60205999132796\\
41	-1.60205999132796\\
42	-1.60205999132796\\
43	-1.60205999132796\\
44	-1.60205999132796\\
45	-1.60205999132796\\
46	-1.60205999132796\\
47	-1.60205999132796\\
48	-1.60205999132796\\
49	-1.60205999132796\\
50	-1.60205999132796\\
51	-1.60205999132796\\
52	-1.60205999132796\\
53	-1.60205999132796\\
54	-1.60205999132796\\
55	-1.60205999132796\\
56	-1.60205999132796\\
57	-1.60205999132796\\
58	-1.60205999132796\\
59	-1.60205999132796\\
60	-1.60205999132796\\
61	-1.60205999132796\\
62	-1.60205999132796\\
63	-1.60205999132796\\
64	-1.60205999132796\\
65	-1.60205999132796\\
66	-1.60205999132796\\
67	-1.60205999132796\\
68	-1.60205999132796\\
69	-1.60205999132796\\
70	-1.60205999132796\\
71	-1.60205999132796\\
72	-1.60205999132796\\
73	-1.60205999132796\\
74	-1.60205999132796\\
75	-1.60205999132796\\
76	-1.60205999132796\\
77	-1.60205999132796\\
78	-1.60205999132796\\
79	-1.60205999132796\\
80	-1.60205999132796\\
81	-1.60205999132796\\
82	-1.60205999132796\\
83	-1.60205999132796\\
84	-1.60205999132796\\
85	-1.60205999132796\\
86	-1.60205999132796\\
87	-1.60205999132796\\
88	-1.60205999132796\\
89	-1.60205999132796\\
90	-1.60205999132796\\
91	-1.60205999132796\\
92	-1.60205999132796\\
93	-1.60205999132796\\
94	-1.60205999132796\\
95	-1.60205999132796\\
96	-1.60205999132796\\
97	-1.60205999132796\\
98	-1.60205999132796\\
99	-1.60205999132796\\
100	-1.60205999132796\\
101	-1.60205999132796\\
102	-1.60205999132796\\
103	-1.60205999132796\\
104	-1.60205999132796\\
105	-1.60205999132796\\
106	-1.60205999132796\\
107	-1.60205999132796\\
108	-1.60205999132796\\
109	-1.60205999132796\\
110	-1.60205999132796\\
111	-1.60205999132796\\
112	-1.60205999132796\\
113	-1.60205999132796\\
114	-1.60205999132796\\
115	-1.60205999132796\\
116	-1.60205999132796\\
117	-1.60205999132796\\
118	-1.60205999132796\\
119	-1.60205999132796\\
120	-1.60205999132796\\
121	-1.60205999132796\\
122	-1.60205999132796\\
123	-1.60205999132796\\
124	-1.60205999132796\\
125	-1.60205999132796\\
126	-1.60205999132796\\
127	-1.60205999132796\\
128	-1.60205999132796\\
129	-1.60205999132796\\
130	-1.60205999132796\\
131	-1.60205999132796\\
132	-1.60205999132796\\
133	-1.60205999132796\\
134	-1.60205999132796\\
135	-1.60205999132796\\
136	-1.60205999132796\\
137	-1.60205999132796\\
138	-1.60205999132796\\
139	-1.60205999132796\\
140	-1.60205999132796\\
141	-1.60205999132796\\
142	-1.60205999132796\\
143	-1.60205999132796\\
144	-1.60205999132796\\
145	-1.60205999132796\\
146	-1.60205999132796\\
147	-1.60205999132796\\
148	-1.60205999132796\\
149	-1.60205999132796\\
150	-1.60205999132796\\
151	-1.60205999132796\\
152	-1.60205999132796\\
153	-1.60205999132796\\
154	-1.60205999132796\\
155	-1.60205999132796\\
156	-1.60205999132796\\
157	-1.60205999132796\\
158	-1.60205999132796\\
159	-1.60205999132796\\
160	-1.60205999132796\\
161	-1.60205999132796\\
162	-1.60205999132796\\
163	-1.60205999132796\\
164	-1.60205999132796\\
165	-1.60205999132796\\
166	-1.60205999132796\\
167	-1.60205999132796\\
168	-1.60205999132796\\
169	-1.60205999132796\\
170	-1.60205999132796\\
171	-1.60205999132796\\
172	-1.60205999132796\\
173	-1.60205999132796\\
174	-1.60205999132796\\
175	-1.60205999132796\\
176	-1.60205999132796\\
177	-1.60205999132796\\
178	-1.60205999132796\\
179	-1.60205999132796\\
180	-1.60205999132796\\
181	-1.60205999132796\\
182	-1.60205999132796\\
183	-1.60205999132796\\
184	-1.60205999132796\\
185	-1.60205999132796\\
186	-1.60205999132796\\
187	-1.60205999132796\\
188	-1.60205999132796\\
189	-1.60205999132796\\
190	-1.60205999132796\\
191	-1.60205999132796\\
192	-1.60205999132796\\
193	-1.60205999132796\\
};
\end{axis}
\end{tikzpicture}%
\caption{The autocorrelation of Model~C residuals (full line) and white noise (dashed line).}
\label{fig:modelC:residuals}
\end{figure}

\subsection{Model~D (linear fourth order)}
\subsubsection{Model formulation}
As visible from \citefig{fig:model:circuitD} and in comparison with Model~C, an additional RC branch further decouples between freezer thermal masses. The complete formulation and the model residuals analysis are skipped because, as shown in the following paragraph, the model extension is not significant.

\begin{figure}[ht]
\centering
\small
\begin{tikzpicture}[scale=2.54]
\ifx\dpiclw\undefined\newdimen\dpiclw\fi
\global\def\dpicdraw{\draw[line width=\dpiclw]}
\global\def\dpicstop{;}
\dpiclw=0.8bp
\dpiclw=0.8bp
\dpiclw=0.85bp
\dpicdraw (0,0.8)
 --(0,0.525)\dpicstop
\dpicdraw (0,0.4) circle (0.049213in)\dpicstop
\dpicdraw (0,0.49375)
 --(0,0.4125)\dpicstop
\filldraw[line width=0bp](-0.026562,0.4125)
 --(0,0.30625)
 --(0.026563,0.4125) --cycle
\dpicstop
\dpicdraw (0,0.275)
 --(0,0)\dpicstop
\draw (0.125,0.4) node[right=-1.5bp]{$ Q$};
\dpicdraw (0,0.8)
 --(0.4,0.8)\dpicstop
\draw (0.4,0.95) node{$V_e$};
\dpicdraw[fill=black](0.4,0.8) circle (0.007874in)\dpicstop
\dpicdraw (0.4,0.8)
 --(0.4,0.425)\dpicstop
\dpicdraw (0.316667,0.425)
 --(0.483333,0.425)\dpicstop
\dpicdraw (0.316667,0.375)
 --(0.483333,0.375)\dpicstop
\dpicdraw (0.4,0.375)
 --(0.4,0)\dpicstop
\draw (0.483333,0.4) node[right=-1.5bp]{$ C_e$};
\dpicdraw[fill=black](0.4,0) circle (0.007874in)\dpicstop
\dpicdraw (0.4,0.8)
 --(0.535,0.8)
 --(0.555833,0.841667)
 --(0.5975,0.758333)
 --(0.639167,0.841667)
 --(0.680833,0.758333)
 --(0.7225,0.841667)
 --(0.764167,0.758333)
 --(0.785,0.8)
 --(0.92,0.8)\dpicstop
\draw (0.66,0.841667) node[above=-1.5bp]{$ R_{e}$};
\draw (0.92,0.95) node{$V_a$};
\dpicdraw[fill=black](0.92,0.8) circle (0.007874in)\dpicstop
\dpicdraw (0.92,0.8)
 --(0.92,0.425)\dpicstop
\dpicdraw (0.836667,0.425)
 --(1.003333,0.425)\dpicstop
\dpicdraw (0.836667,0.375)
 --(1.003333,0.375)\dpicstop
\dpicdraw (0.92,0.375)
 --(0.92,0)\dpicstop
\draw (1.003333,0.4) node[right=-1.5bp]{$ C_a$};
\dpicdraw[fill=black](0.92,0) circle (0.007874in)\dpicstop
\dpicdraw (0.92,0.8)
 --(1.055,0.8)
 --(1.075833,0.841667)
 --(1.1175,0.758333)
 --(1.159167,0.841667)
 --(1.200833,0.758333)
 --(1.2425,0.841667)
 --(1.284167,0.758333)
 --(1.305,0.8)
 --(1.44,0.8)\dpicstop
\draw (1.18,0.841667) node[above=-1.5bp]{$ R_{a}$};
\draw (1.44,0.95) node{$V_w$};
\dpicdraw[fill=black](1.44,0.8) circle (0.007874in)\dpicstop
\dpicdraw (1.44,0.8)
 --(1.44,0.425)\dpicstop
\dpicdraw (1.356667,0.425)
 --(1.523333,0.425)\dpicstop
\dpicdraw (1.356667,0.375)
 --(1.523333,0.375)\dpicstop
\dpicdraw (1.44,0.375)
 --(1.44,0)\dpicstop
\draw (1.523333,0.4) node[right=-1.5bp]{$ C_w$};
\dpicdraw[fill=black](1.44,0) circle (0.007874in)\dpicstop
\dpicdraw (1.44,0.8)
 --(1.575,0.8)
 --(1.595833,0.841667)
 --(1.6375,0.758333)
 --(1.679167,0.841667)
 --(1.720833,0.758333)
 --(1.7625,0.841667)
 --(1.804167,0.758333)
 --(1.825,0.8)
 --(1.96,0.8)\dpicstop
\draw (1.7,0.841667) node[above=-1.5bp]{$ R_{w}$};
\draw (1.96,0.95) node{$V_f$};
\dpicdraw[fill=black](1.96,0.8) circle (0.007874in)\dpicstop
\dpicdraw (1.96,0.8)
 --(1.96,0.425)\dpicstop
\dpicdraw (1.876667,0.425)
 --(2.043333,0.425)\dpicstop
\dpicdraw (1.876667,0.375)
 --(2.043333,0.375)\dpicstop
\dpicdraw (1.96,0.375)
 --(1.96,0)\dpicstop
\draw (2.043333,0.4) node[right=-1.5bp]{$ C_f$};
\dpicdraw[fill=black](1.96,0) circle (0.007874in)\dpicstop
\dpicdraw (1.96,0.8)
 --(2.095,0.8)
 --(2.115833,0.841667)
 --(2.1575,0.758333)
 --(2.199167,0.841667)
 --(2.240833,0.758333)
 --(2.2825,0.841667)
 --(2.324167,0.758333)
 --(2.345,0.8)
 --(2.48,0.8)\dpicstop
\draw (2.22,0.841667) node[above=-1.5bp]{$ R_{f}$};
\dpicdraw (2.48,0)
 --(2.48,0.275)\dpicstop
\dpicdraw (2.48,0.4) circle (0.049213in)\dpicstop
\draw (2.48,0.3375) node{$_-$};
\draw (2.48,0.4625) node{$_+$};
\dpicdraw (2.48,0.525)
 --(2.48,0.8)\dpicstop
\draw (2.605,0.4) node[right=-1.5bp]{$ T_r$};
\dpicdraw (2.48,0)
 --(0,0)\dpicstop
\end{tikzpicture}
\caption{Model~D TEC.
}\label{fig:model:circuitD}
\end{figure}
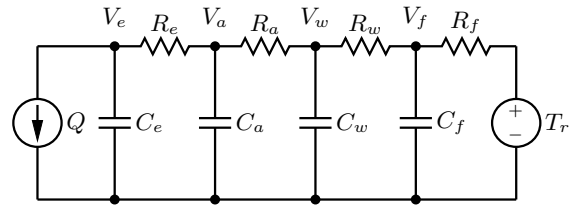

\subsubsection{Model validation}
The deviance of the current model extension is $D=1.0$, that produces a p-value of 61\%. As the p-value is well above the 5\% threshold, the null hypothesis cannot be rejected and the model extension is not valid.

\subsection{Model~E (nonlinear third order)}
\subsubsection{Model formulation}
As known, the \COP\ of an ideal refrigeration cycle is given by the reversed Carnot cycle formula:
\begin{align}
  \COP_\text{ideal}(T_H, T_C) =  \dfrac{T_C + 273}{T_H - T_C} 
  \label{eq:model:rcc},
\end{align}
where $T_H$ and $T_C$ are the temperatures in \SI{}{\celsius} of the hot and cold heat reservoir, respectively.
In other words, the refrigeration cycle ability to extract heat depends on the temperature difference with the exterior. 
This effect was not considered in the previous linear models, where the $\COP$ was modelled as a constant coefficient.
Model~E is a third order model as Model~C where the heat extracted from the freezer chamber is described by:
\begin{align}
  Q = P \cdot \eta \cdot \COP_\text{ideal}(T_r, V_e)\label{},
\end{align}
where $\eta$ can be regarded to as the efficiency of the implemented refrigeration cycle with respect to the ideal case, and the hot and cold heat reservoirs are approximated with the room and freezer cold side temperature. 
With reference to the stochastic state space representation \eqref{eq:ssd:1}-\eqref{eq:ssd:2}, the complete formulation of Model~E is as:
\begingroup
\setlength{\arraycolsep}{3pt}
\begin{align}
\boldsymbol{x}^T &= \begin{bmatrix} V_e & V_a & V_w \end{bmatrix} \label{eq:modelEi}\\
\boldsymbol{u}^T &= \begin{bmatrix} T_r & P\dfrac{V_e + 273}{T_r - V_e} \end{bmatrix}\label{eq:nlinearmodel}\\
A &= \begin{bmatrix}
      \frac{-1}{C_e R_e} &  \frac{1}{C_e R_e} &  0 \\
      \frac{1}{C_a R_e}  & \frac{-1}{C_a R_a}  + \frac{-1}{C_a R_e} &  \frac{1}{C_a R_a} \\
      0 & \frac{1}{C_w R_a} &  \frac{-1}{C_w R_w}  +  \frac{-1}{C_w R_a}
     \end{bmatrix}\\
B &= \begin{bmatrix}
      0 & -\frac{\eta}{C_e} \\
      0 & 0 \\
      \frac{1}{C_w R_w} & 0
     \end{bmatrix} \\
W &= \text{diag}(w_0, w_1, w_2)\\
C &= \begin{bmatrix} 0 & 1 & 0 \end{bmatrix} \label{eq:modelEf}
\end{align}
\endgroup
that is nonlinear as the input \eqref{eq:nlinearmodel} depends on a component of the state vector.

\subsubsection{Model identification and validation}
The autocorrelation of the model residuals is shown in \citefig{fig:modelE:residuals} and has a similar behavior as for Model~C. Again, only a few components of the autocorrelation function are above the dotted line, thus indicating a good overall capacity of Model~E to describe the dynamics contained in the training data set.
The deviance test is not computed because Model~E is not an extension of the previous, rather it relies on a different mathematical description of the refrigeration process. Nevertheless, it is worth noting than Model~E has a better fitting than Model~C because, although the two have same number of parameters, Model~E achieves a larger log-likelihood value.

\begin{figure}[!ht]
\centering
\scriptsize
%
\begin{tikzpicture}

\begin{axis}[%
width=\matlabfigurewidth,
height=0.362327188940092\matlabfigurewidth,
scale only axis,
xmin=0,
xmax=140,
xlabel={Lag},
ymin=-6,
ymax=0,
ylabel={$\log_{10}(|\text{Residuals ACF}|)$}
]
\addplot [color=black,solid,mark=o,mark options={solid},forget plot]
  table[row sep=crcr]{%
1	-4.82163733276644e-17\\
2	-1.54739679712004\\
3	-1.53136703635349\\
4	-2.32622939594707\\
5	-2.29213343476291\\
6	-3.44000315288554\\
7	-2.01125904459996\\
8	-2.25841396257457\\
9	-2.85329168936883\\
10	-2.13531484720006\\
11	-2.47110905877566\\
12	-2.14570818828435\\
13	-1.89702813929774\\
14	-3.17041883089626\\
15	-2.2421594574487\\
16	-2.80972551236114\\
17	-2.2333788363624\\
18	-3.22538930800578\\
19	-1.89925967776633\\
20	-2.19977496464897\\
21	-3.58460471868479\\
22	-1.88790350633069\\
23	-3.57879430207582\\
24	-1.88566279607484\\
25	-1.95488047408999\\
26	-2.038629575871\\
27	-2.29748158082822\\
28	-2.08537394557535\\
29	-2.64671531653708\\
30	-1.95355702968177\\
31	-1.94020871782769\\
32	-1.8559923936956\\
33	-2.04441248676929\\
34	-2.13026037723362\\
35	-2.33341396854227\\
36	-2.36748381248097\\
37	-2.61441703062005\\
38	-1.86229422814812\\
39	-2.52644034188948\\
40	-1.87215092661483\\
41	-2.7932366564778\\
42	-1.95144846839893\\
43	-2.19126343526445\\
44	-2.26816084540355\\
45	-2.78559640108381\\
46	-2.71478831275737\\
47	-2.60944641137999\\
48	-2.56451843762729\\
49	-2.3154275291012\\
50	-2.65395662636074\\
51	-4.10010004058913\\
52	-2.00467497733157\\
53	-2.34281729259424\\
54	-2.12373239808773\\
55	-2.17162673789456\\
56	-2.51260631708802\\
57	-2.07717212193238\\
58	-1.95036833165656\\
59	-2.26727077747127\\
60	-2.3131659630621\\
61	-2.24367214382295\\
62	-2.66697949307457\\
63	-2.45230801783411\\
64	-2.20108390297175\\
65	-2.25447009768131\\
66	-3.01703764059246\\
67	-2.30605333278173\\
68	-2.24398366218052\\
69	-2.51918694615792\\
70	-2.26715728110617\\
71	-2.2011502190084\\
72	-2.32509055314618\\
73	-1.75709474688142\\
74	-1.86963913841068\\
75	-2.16651933008054\\
76	-2.2614802860743\\
77	-2.04235009382387\\
78	-2.39654366959725\\
79	-2.86862633654029\\
80	-2.08747504837421\\
81	-1.96245388886255\\
82	-2.03761853876131\\
83	-2.29510257146153\\
84	-2.79328258152207\\
85	-2.21276557929522\\
86	-3.42538977509661\\
87	-2.2211249797169\\
88	-2.38963104140217\\
89	-2.29712634707141\\
90	-2.14553222500171\\
91	-2.61387495063619\\
92	-2.82550495460648\\
93	-2.94416501342997\\
94	-2.32964950133267\\
95	-2.10747094102006\\
96	-2.5590980636454\\
97	-1.87631290226273\\
98	-2.54642725122126\\
99	-2.39590891158968\\
100	-2.26705626909773\\
101	-2.35661473810515\\
102	-1.84697189427265\\
103	-2.06412094081286\\
104	-2.12588418315165\\
105	-2.01641143821376\\
106	-1.87193917667573\\
107	-2.32476227330675\\
108	-2.65026239779513\\
109	-2.0648297922811\\
110	-2.69110953455572\\
111	-1.90538197898589\\
112	-2.21539039864889\\
113	-1.71326267097653\\
114	-3.00586415139417\\
115	-1.79064581851774\\
116	-3.23939482616701\\
117	-2.07055263487801\\
118	-2.91124120653421\\
119	-1.98018213811458\\
120	-2.3992430596994\\
121	-2.06455651107418\\
122	-2.9777430303358\\
123	-2.05294516582299\\
124	-2.22781749314857\\
125	-1.59822500617904\\
126	-1.95746937536124\\
127	-1.88986473646367\\
128	-2.28604676904113\\
129	-2.29209611691614\\
130	-3.49308836419735\\
131	-2.35337375719013\\
132	-2.39441515695696\\
133	-2.35016035616787\\
134	-2.09566983874458\\
135	-2.46796961048228\\
136	-1.61687367319632\\
137	-2.0861265558913\\
138	-1.77932586403913\\
139	-2.13419369483177\\
140	-2.25933491047315\\
141	-3.23207163309208\\
142	-2.39121877064026\\
143	-1.98224997771442\\
144	-2.78473211471254\\
145	-1.93965987729865\\
146	-2.0164073052937\\
147	-1.86126375621357\\
148	-1.8170409774411\\
149	-1.78974321657072\\
150	-2.6277690510186\\
151	-2.15800247359201\\
152	-2.8290909386128\\
153	-2.42041234130967\\
154	-1.87523028333789\\
155	-3.24441772698342\\
156	-2.39866114454093\\
157	-3.56372213035329\\
158	-2.07274311004305\\
159	-1.89608853628689\\
160	-3.0049407853619\\
161	-1.90951654234952\\
162	-2.27521647673118\\
163	-2.7300561466948\\
164	-2.16296826028469\\
165	-2.70609354928549\\
166	-2.56638175679745\\
167	-2.37305970370719\\
168	-2.23289751007881\\
169	-2.27273716248893\\
170	-2.60506712156746\\
171	-2.10081987769781\\
172	-2.52603600436802\\
173	-3.52890028424453\\
174	-2.55523461366795\\
175	-1.9318162637446\\
176	-1.76643981407442\\
177	-2.14851466954634\\
178	-2.6283951051666\\
179	-2.57976075224858\\
180	-3.05912632551987\\
181	-2.33026224131653\\
182	-2.59584872086093\\
183	-2.66467690319384\\
184	-3.86653189441316\\
185	-2.57182007229317\\
186	-2.06915046446152\\
187	-2.46504919905499\\
188	-2.5401941935835\\
189	-3.21176837680919\\
190	-1.88228037634316\\
191	-2.72663351896764\\
192	-2.23855255027095\\
193	-1.87139223708798\\
};
\addplot [color=red,dashed,line width=1.5pt,forget plot]
  table[row sep=crcr]{%
1	-1.60205999132796\\
2	-1.60205999132796\\
3	-1.60205999132796\\
4	-1.60205999132796\\
5	-1.60205999132796\\
6	-1.60205999132796\\
7	-1.60205999132796\\
8	-1.60205999132796\\
9	-1.60205999132796\\
10	-1.60205999132796\\
11	-1.60205999132796\\
12	-1.60205999132796\\
13	-1.60205999132796\\
14	-1.60205999132796\\
15	-1.60205999132796\\
16	-1.60205999132796\\
17	-1.60205999132796\\
18	-1.60205999132796\\
19	-1.60205999132796\\
20	-1.60205999132796\\
21	-1.60205999132796\\
22	-1.60205999132796\\
23	-1.60205999132796\\
24	-1.60205999132796\\
25	-1.60205999132796\\
26	-1.60205999132796\\
27	-1.60205999132796\\
28	-1.60205999132796\\
29	-1.60205999132796\\
30	-1.60205999132796\\
31	-1.60205999132796\\
32	-1.60205999132796\\
33	-1.60205999132796\\
34	-1.60205999132796\\
35	-1.60205999132796\\
36	-1.60205999132796\\
37	-1.60205999132796\\
38	-1.60205999132796\\
39	-1.60205999132796\\
40	-1.60205999132796\\
41	-1.60205999132796\\
42	-1.60205999132796\\
43	-1.60205999132796\\
44	-1.60205999132796\\
45	-1.60205999132796\\
46	-1.60205999132796\\
47	-1.60205999132796\\
48	-1.60205999132796\\
49	-1.60205999132796\\
50	-1.60205999132796\\
51	-1.60205999132796\\
52	-1.60205999132796\\
53	-1.60205999132796\\
54	-1.60205999132796\\
55	-1.60205999132796\\
56	-1.60205999132796\\
57	-1.60205999132796\\
58	-1.60205999132796\\
59	-1.60205999132796\\
60	-1.60205999132796\\
61	-1.60205999132796\\
62	-1.60205999132796\\
63	-1.60205999132796\\
64	-1.60205999132796\\
65	-1.60205999132796\\
66	-1.60205999132796\\
67	-1.60205999132796\\
68	-1.60205999132796\\
69	-1.60205999132796\\
70	-1.60205999132796\\
71	-1.60205999132796\\
72	-1.60205999132796\\
73	-1.60205999132796\\
74	-1.60205999132796\\
75	-1.60205999132796\\
76	-1.60205999132796\\
77	-1.60205999132796\\
78	-1.60205999132796\\
79	-1.60205999132796\\
80	-1.60205999132796\\
81	-1.60205999132796\\
82	-1.60205999132796\\
83	-1.60205999132796\\
84	-1.60205999132796\\
85	-1.60205999132796\\
86	-1.60205999132796\\
87	-1.60205999132796\\
88	-1.60205999132796\\
89	-1.60205999132796\\
90	-1.60205999132796\\
91	-1.60205999132796\\
92	-1.60205999132796\\
93	-1.60205999132796\\
94	-1.60205999132796\\
95	-1.60205999132796\\
96	-1.60205999132796\\
97	-1.60205999132796\\
98	-1.60205999132796\\
99	-1.60205999132796\\
100	-1.60205999132796\\
101	-1.60205999132796\\
102	-1.60205999132796\\
103	-1.60205999132796\\
104	-1.60205999132796\\
105	-1.60205999132796\\
106	-1.60205999132796\\
107	-1.60205999132796\\
108	-1.60205999132796\\
109	-1.60205999132796\\
110	-1.60205999132796\\
111	-1.60205999132796\\
112	-1.60205999132796\\
113	-1.60205999132796\\
114	-1.60205999132796\\
115	-1.60205999132796\\
116	-1.60205999132796\\
117	-1.60205999132796\\
118	-1.60205999132796\\
119	-1.60205999132796\\
120	-1.60205999132796\\
121	-1.60205999132796\\
122	-1.60205999132796\\
123	-1.60205999132796\\
124	-1.60205999132796\\
125	-1.60205999132796\\
126	-1.60205999132796\\
127	-1.60205999132796\\
128	-1.60205999132796\\
129	-1.60205999132796\\
130	-1.60205999132796\\
131	-1.60205999132796\\
132	-1.60205999132796\\
133	-1.60205999132796\\
134	-1.60205999132796\\
135	-1.60205999132796\\
136	-1.60205999132796\\
137	-1.60205999132796\\
138	-1.60205999132796\\
139	-1.60205999132796\\
140	-1.60205999132796\\
141	-1.60205999132796\\
142	-1.60205999132796\\
143	-1.60205999132796\\
144	-1.60205999132796\\
145	-1.60205999132796\\
146	-1.60205999132796\\
147	-1.60205999132796\\
148	-1.60205999132796\\
149	-1.60205999132796\\
150	-1.60205999132796\\
151	-1.60205999132796\\
152	-1.60205999132796\\
153	-1.60205999132796\\
154	-1.60205999132796\\
155	-1.60205999132796\\
156	-1.60205999132796\\
157	-1.60205999132796\\
158	-1.60205999132796\\
159	-1.60205999132796\\
160	-1.60205999132796\\
161	-1.60205999132796\\
162	-1.60205999132796\\
163	-1.60205999132796\\
164	-1.60205999132796\\
165	-1.60205999132796\\
166	-1.60205999132796\\
167	-1.60205999132796\\
168	-1.60205999132796\\
169	-1.60205999132796\\
170	-1.60205999132796\\
171	-1.60205999132796\\
172	-1.60205999132796\\
173	-1.60205999132796\\
174	-1.60205999132796\\
175	-1.60205999132796\\
176	-1.60205999132796\\
177	-1.60205999132796\\
178	-1.60205999132796\\
179	-1.60205999132796\\
180	-1.60205999132796\\
181	-1.60205999132796\\
182	-1.60205999132796\\
183	-1.60205999132796\\
184	-1.60205999132796\\
185	-1.60205999132796\\
186	-1.60205999132796\\
187	-1.60205999132796\\
188	-1.60205999132796\\
189	-1.60205999132796\\
190	-1.60205999132796\\
191	-1.60205999132796\\
192	-1.60205999132796\\
193	-1.60205999132796\\
};
\end{axis}
\end{tikzpicture}%
\caption{The autocorrelation of Model~E residuals (full line) and white noise (dashed line).}
\label{fig:modelE:residuals}
\end{figure}
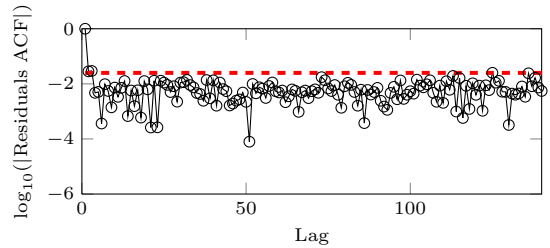


\section{An empirical estimation of the freezer components thermal characteristics}\label{sec:whitebox}
In this section, we perform a purely empirical estimation of the physical values of the main freezer components.
This process should not be meant as a replacement of the previous grey-box modeling methodology, that is in fact a more general and powerful tool as parameters are estimated from measurements, regardless of unknown or approximative information on the physical characteristics of the freezer components. 
Moreover, grey-box models have the advantage of being tuned on the specific device to model, a degree of freedom that is not achievable by first principles based models.
Nevertheless, the analysis proposed in this section is of importance to verify whether the previously identified parameters are of reasonable order of magnitude and if the circuit components absorbed those dynamics they were designed for.
The estimated  physical characteristics of the main freezer components are in Table \ref{tab:rough:pars}.
The thermal capacities are calculated as
\begin{align}
 C = M \cdot c,
\end{align}
where $M$, $c$ respectively denote a mass (calculated as a volume times the density of the component) and specific heat capacity. Volumes are estimated by measuring the size of the components. In the case of a non-accessible part, the size was reasonably guessed. 
The thermal resistance of the isolation layer is computed as
\begin{align}
 R = \frac{1}{\lambda} \cdot \frac{t}{S}
\end{align}
where $\lambda, t, S$ respectively denote the thermal conductivity, thickness and lateral surface of the freezer envelope. 
Discrepancies can be noted when comparing the values in Table \ref{tab:rough:pars} with the fitted model parameters. For example, in the case of Model~E, the value of $C_a$ is approximately 2 orders of magnitude larger than the empirically calculated thermal capacity of the air, and vice-versa in the case of $C_e$ and the thermal capacity of the heat exchanger. This indicates that the capacitors of Model~E did not absorbed those dynamics for which it was originally thought. Rather, $C_a$ is a lumped description of the thermal mass of several freezer parts.
In spite of this, the total values of the best fitting models thermal capacity and resistance are with same order of magnitude as the global empirical, an indication that the found values of the models parameters are globally meaningful. 

\begin{table}[!ht]
\centering
{ \scriptsize
\renewcommand{\arraystretch}{1.2}
\caption{Physical properties and empirically estimated characteristics of the freezer main components.}\label{tab:rough:pars}
\begin{tabular}{C{1.8cm}|C{1.15cm}|C{1.7cm}|C{1.4cm}}
\hline
Component & Chamber & Isolation Layer & Heat Exchanger \\
\hline
Material & Air & Polyurethane foam \cite{polyurethane} & Aluminium \\
\hline
Specific heat capacity $c$ (\SI{}{\joule\per\kilo\gram\per\kelvin}) & 1000 & 1500 & 897 \\
\hline
Total mass $M$ (\SI{}{\kilo\gram}) & 0.5 & 8 & 10 \\
\hline
Thermal capacity $C$ (\SI{}{\joule\per\kelvin}) & \SI{5e2}{} & \SI{5e3}{} & \SI{1e4}{} \\
\hline
Thermal conductivity $\lambda$ (\SI{}{\watt\per\meter\per\kelvin}) & -- & 0.025 & -- \\
\hline
Thickness $t$ (m) & -- & 0.08 & -- \\
\hline
Lateral surface $S$ (\SI{}{\meter\squared}) & -- & 2.4 & -- \\
\hline
Thermal resistance $R$ (\SI{}{\kelvin\per\watt}) & -- & 1.3 & -- \\
\hline
\end{tabular}
}
\end{table}


\section{Models Performance Assessment}\label{sec:modelsperformance}
\citetable{tab:comp:comp} shows, for each identified model, the mean and standard deviation of the model residuals derived from 20-minute-ahead predictions using a validation data set. 
The best performing model (\ie\ smallest residuals bias and standard deviation) is Model~E, while Model~A is the worst, thus overall confirming the model identification results. Moreover, Model~D does not show better performance than Model~C, therefore validating the outcome of the previous inference analysis according to which the extension to a fourth order linear model was not statistically significant.

\begin{table}[ht]
\scriptsize
\renewcommand{\arraystretch}{1.4}
\caption{Model residuals statistics for 20-minute ahead predictions with a PRBS validation data set.}\label{tab:comp:comp}
\centering
\begin{tabular}{c c c}
\hline
Model & $\bar{e}$ [\SI{}{\celsius}] & $\sigma_e$ [\SI{}{\celsius}]\\
\midrule
Model~A & 0.593  & 2.8 \\
Model~B & 0.227 & 0.91 \\
Model~C & 0.106 & 0.60\\
Model~D &  0.145  & 0.74 \\
Model~E &  0.044 & 0.45 \\
\hline
\end{tabular}
\end{table}

\citetable{tab:comp:comp_smaller} shows the same statistical analysis as in the previous table, but performed with a validation data set measured under conventional thermostatic control for the purpose of highlighting the prediction performance of the models during conventional operation.  Overall, the prediction performance of the models in increased. This could be explained by the fact that under thermostatic regime, the temperature of the freezer varies in a smaller range than in the PRBS case, therefore possible nonlinear effects due to temperature variations not explained in the models are reduced.

\begin{table}[ht]
\scriptsize
\caption{Model residuals statistics for 20-minute ahead predictions under conventional thermostatic regime.}\label{tab:comp:comp_smaller}
\renewcommand{\arraystretch}{1.4}
\centering
\begin{tabular}{c c c c c}
\hline
Model & $\bar{e}$ [\SI{}{\celsius}] & $\sigma_e$ [\SI{}{\celsius}]\\
\midrule
Model~B &  0.15  & 0.73 \\
Model~C &  0.056  & 0.29 \\
Model~E &  0.023 & 0.27 \\
\hline
\end{tabular}
\end{table}

\section{Optimizing the power consumption of a freezer using model predictive control}\label{sec:application}
\subsection{Introduction and objective}
A paradigm often advocated in the existing literature to increase the proportion of electricity production from renewables is to restore an adequate level of controllability by giving the possibility of shifting the consumption of DSRs \cite{Herter20101561, Zarnikau20101536, Cappers20101526, keane2011demand, Hedegaard2012356}. 
Electric heating systems, water heaters and refrigeration units are all loads that, although with different levels of flexibility, can be controlled to temporarily defer the consumption thanks to the their thermal mass. 
Among several algorithms for shifting the consumption of flexible demand, MPC comes to prominence as a method to achieve the non disruptive controllability of individual DSRs through a consumption incentive signal, like for example a dynamic electricity price \cite{halvgaard2012economic, 6019347} or according to the availability of renewable production, as done in \cite{FASO_ISGT2013, callaway2009tapping}. 
MPC consists in determining the electrical power consumption trajectory of a DSR that minimizes a given penalty function (like the total cost of the operation) while obeying to consumer comfort and operational constraints by implementing a DSR prediction model.

In this section, we describe an experimental application of MPC to achieve a shift in the consumption of the previously described instrumented freezer.
The MPC experiments are carried out implementing several of the freezer presented in \citesec{sec:models}, with the main objective being to assess their performance in a practical application.

\subsection{MPC general formulation}\label{sec:mpcgeneralformulation}
The MPC strategy is formulated and actuated at discrete time intervals of duration $d$. The index $i$ denotes the rolling time interval, while $k$ is a generic discrete time index that rolls over the prediction horizon. The freezer temperature prediction models, which were formulated in continuous time, are discretized as shown in \ref{app:mpcdev}. 
The objective of the MPC formulation is to determine the sequence $\boldsymbol{P}^o_i=\begin{bmatrix} P^o_i & P^o_{i+1} & \cdots & P^o_{i+N} \end{bmatrix} \in \mathbb{R}^{N+1}$, \ie\ the freezer power consumption from the current time instant and for the next $N$. 
The formulation consists of the following optimization problem:
\begin{align}
\boldsymbol{P}^o_i = \argmin{\boldsymbol{P}_i \in \mathcal{P}} \sum_{k=i}^{i+N} P_{k}\cdot c_{k} \label{mpc:cost0}
\end{align}
subject to:
\begin{align}
 & \overline{T}_{k+1} = f(T_{i}, P_k, T_{r,k}), & k=i,\dots,i+N \\
 & T_\text{min} \le \overline{T}_{k+1} \le T_\text{max}, & k=i,\dots,i+N \label{eq:Tconstr}\\
 & 0 \le P_{k} \le P_\text{max}, & k=i,\dots,i+N, \label{eq:Pconstr}
\end{align}
where $f(\cdot)$ denotes a discrete time freezer model, $\overline{T}$ is the expected value of the freezer temperature prediction, $T_i$ is the freezer current temperature (from measurements), $T_{r}$ is the room temperature, $T_\text{min}$, $T_\text{max}$ define the range where the freezer temperature is allowed, and $P_\text{max}$ is the maximum power consumption of the freezer. As shown in \ref{app:mpcdev}, freezer linear models lead to convex optimization problems because the inequality in ~\eqref{eq:Tconstr} can be written as a linear function of the decision vector. 
As known, convexity is an appealing property for optimization problems because it is a sufficient condition for the uniqueness of the solution. 
Said otherwise, if a solution to the problem exists, it is the global optimum. Moreover, there exist efficient algorithms to solve convex optimization problems. On the contrary, the optimization problem of the MPC with the nonlinear freezer model is nonconvex. This aspect will be further addressed when presenting the experimental results.
Said in words, the optimization problem in \eqref{mpc:cost0}-\eqref{eq:Pconstr} seeks for the freezer power consumption trajectory that minimizes the penalty function while respecting the following operational constraints:
\begin{itemize}
 \item keeping an adequate temperature level to preserve food quality according to the consumer preferences. According to food storage regulation  \cite{fdafoodregulation, ukfoodregulation}, freezer temperature should be regulated at \SI{-18}{\celsius}, although temperatures in the range from \SI{-22}{} to \SI{-28}{\celsius} are indicated to achieve longer storage periods. In this case, the temperature bounds are chosen as $T_\text{min}=\SI{-27}{\celsius}$ and $T_\text{max}=\SI{-18}{\celsius}$.
  
 \item the freezer power consumption during the on phase of the compressor is modelled as a constant term, and it is limited by the maximum power absorption of the freezer, roughly \SI{68}{\watt}. We recall from \citesec{sec:expSetup} that the compressor can be only on-off controlled. Therefore, we use the following procedure to turn the real scalar power consumption set-points of \eqref{mpc:cost0} into a sequence of on/off pulses of period $d$:
 \begin{align}
  \tau_{\text{on}, k} = \frac{P_k}{P_\text{max}} d, \ \ k=i,\cdots,i+N \label{eq:pwm}
 \end{align}
 where $\tau_{\text{on}}$ is the duration in second of the \emph{on} phase. In other words, Eq.~\eqref{eq:pwm} performs a pulse width modulation (PWM) to translate the MPC power set-point into an on-off signal that, over the period $d$, delivers an equal amount of electricity. The ratio $\tau_{\text{on}, k}/d$ is called duty cycle. This formulation is convenient because it does not require to formulate the optimization problem as a mixed integer programming problem.
\end{itemize}
The sequence $c_i, \dots, c_N$ in \eqref{mpc:cost0} is a virtual electricity price that must be known in advance for the whole length of the optimization period. In this case, it is chosen as a step signal to allow for evaluating the amount of freezer consumption that the MPC can shift in view of an increase in the electricity price. The constant coefficient $N$ in \eqref{mpc:cost0} determines the length of the look-ahead horizon. 
The electricity that can be virtually stored in the freezer is given by the amount of consumption that is shifted while respecting the temperature constraints. 
Therefore, $N$ should be chosen larger than the time that the freezer temperature takes to go from the upper to the lower bound or, in other words, larger than the time required to saturate the storage capacity.
As highlighted in \cite[p.~36]{sossan_thesis}, it is worth noting that the formulation \eqref{mpc:cost0}-\eqref{eq:Pconstr} does not react to stepwise decreases of the price signal. In the case such a functionality is of interest, one should considered to implement a quadratic cost function to penalize deviation from a given temperature set-point.

\begin{figure}[ht]
\centering {
\scriptsize
\tikzstyle{decision} = [diamond, draw, fill=blue!0, 
    text width=4.5em, text badly centered, node distance=3cm, inner sep=0pt]
\tikzstyle{block} = [rectangle, draw, fill=blue!10, 
    text width=15em, text centered, rounded corners, minimum height=3em]
\tikzstyle{line} = [draw, -latex']
    
\begin{tikzpicture}[node distance = 1.25cm, auto]
    \node [block, text width=13em] (init) {$i=0$, Initialize model};
    \node [block, below of=init] (cc) {Read the virtual electricity price};
    \node [block, below of=cc] (evaluate) {Read freezer temperature and update model state with KF/EKF};
    \node [block, below of=evaluate] (solve) {Solve the optimization problem in the horizon $i$ to $i+N$};
    \node [block, below of=solve] (apply) {Actuate the first decision of the MPC control law using PWM};
    \node [block, below of=apply] (wait) {Wait for $d$~second};
    
    \node [block, left of=solve, node distance=4cm, text width=5em] (next) {$i=i+1$};

    \path [line] (init) -- (cc);
    \path [line] (cc) -- (evaluate);
    \path [line] (evaluate) -- (solve);
    \path [line] (solve) -- (apply);
    \path [line] (apply) -- (wait);
    \path [line] (wait) -| (next);
    \path [line] (next) |- (cc);
\end{tikzpicture}
}
\caption{Flow chart of the MPC experiment illustrating the receding horizon policy.}\label{fig:flowchart}
\end{figure}
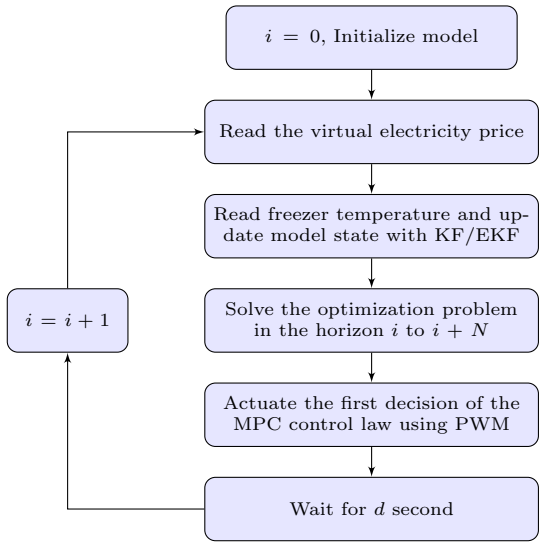

\subsection{On the actuation of the MPC law}
As is usually the case, the MPC action is actuated in a receding horizon way, meaning that, at each time interval, the state vector of the prediction model is updated using the last available measurements from the freezer, the optimization problem is solved and the first portion of the MPC control law is applied. 
A diagram summarizing the complete sequence of events performed during the experiments is shown in \citefig{fig:flowchart}. At the initial stage, the prediction model is initialized with the latest freezer temperature measurement (steady state conditions are assumed). Then, at each iteration of the receding horizon cycle, the virtual electricity price for the next $N$ periods is acquired, as well as the current temperature measurement. The latter information is used to update the state of the prediction model by a using a Kalman filter (KF). The prediction stage of the KF consists in determining the model state and covariance matrix evolution as:
\begin{align}
 \widehat{\boldsymbol{x}}_{k|k-1} &= A_d \widehat{\boldsymbol{x}}_{k-1|k-1} + B_d \boldsymbol{u}_{k-1} \\
 P_{k|k-1} &= A_d P_{k-1|k-1} A_d^T + WW^T \label{eq:kf:sc},
\end{align}
where $A_d$, $B_d$ are the linear discrete state space model matrices (derived in \ref{app:mpcdev}), $C$ is the output vector and $W$ is the noise process matrix determined in the parameters estimation process. The Kalman gain is
\begin{align}
 K = P_{k|k-1}C^T\left( C P_{k|k-1} C^T + v^2\right)^{-1} \label{eq:kf:kg},
\end{align}
where $v$ is the measurement noise (also known from the parameters estimation). Once the new measurement $y_k$ is available, the state prediction is updated as:
\begin{align}
 \widehat{\boldsymbol{x}}_{k|k} &= \widehat{\boldsymbol{x}}_{k|k-1} + K_k(y_k - C\widehat{\boldsymbol{x}}_{k|k-1}) \label{eq:kf:u1} \\
 P_{k|k} &= \left(P_{k|k-1}^{-1} + C^Tv^{-1}C\right)^{-1}. \label{eq:kf:u2}
\end{align}
Instead, in the case of Model~E, the extended  Kalman filter is used. The state prediction is performed as:
\begin{align}
 \widehat{\boldsymbol{x}}_{k|k-1} = f(\widehat{\boldsymbol{x}}_{k|k}, P_i, T_{r, i})
\end{align}
using the discretized version of the nonlinear model in \eqref{eq:modelEi}-\eqref{eq:modelEf}. The state covariance matrix, KF gain and update steps are performed using the equations \eqref{eq:kf:sc}-\eqref{eq:kf:u2} as in the previous case, but $A_d$ is now as:
\begin{align}
 A_d = \dfrac{\partial f}{\partial \boldsymbol x}\bigg\rvert_{x_{k-1|k-1}, u_{k-1}},
\end{align}
namely the first order Taylor expansion of the model with respect to the state vector. Finally from \citefig{fig:flowchart}, once the model is updated, the optimization problem is solved in order to determine the optimal control law $\boldsymbol{P}^o_i$. Then, the decision $P^o_i$ for the current instant of time is extracted and actuated in the PWM sense (\citeeq{eq:pwm}) by regulating the on-off timing of the controllable freezer power plug. A further consideration concerns the implementation of temperature soft constraints in the optimization problem. In fact, in the formulation  \eqref{mpc:cost0}-\eqref{eq:Pconstr}, the freezer temperature is strictly allowed only in a well determined range. If the temperature hard constraint \eqref{eq:Tconstr} is not satisfied (for example at the time instant $i+1$ because unmodelled system dynamics, noise or consumer behavior), the optimization problem is unfeasible, causing a failure of the control system. It is therefore convenient to add to \eqref{mpc:cost0}-\eqref{eq:Pconstr} a sequence of positive slack variables $\boldsymbol{s} \in \mathbb{R}^{N+1}$ to relax the constraints:
\begin{align}
\{\boldsymbol{P}^o_i, \boldsymbol{s}^o_i\}= \argmin{\{\boldsymbol{P}_i,\boldsymbol{s}_i\} \in \Omega} \left(\sum_{k=i}^N P_{k}\cdot c_{k} + b_k \cdot s_k \right) \label{eq:MPC:realcost}
\end{align}
subject to:
\begin{align}
 & \overline{T}_{k+1} = f(T_{i}, P_k, T_{r,k}), & k=i,\dots,i+N \\
 & \overline{T}_{k+1} \le T_\text{max} + s_k, & k=i,\dots,i+N \\
 & \overline{T}_{k+1} \ge T_\text{min} - s_k  & k=i,\dots,i+N \\
 & s_k \ge 0  & k=i,\dots,i+N \\
 & 0 \le P_{k} \le P_\text{max}, & k=i,\dots,i+N, \label{eq:MPC:reale}
\end{align}
where $b_k$ for $ k=i+1,\dots,i+N$ are coefficients that should be chosen much larger than $c_k$.
In this way, deviations from the optimal temperature range are allowed but not convenient because they are strongly penalized in the cost function. 
A final aspect regards the actuation of the PWM signal and the on/off transitions of the freezer, which are of concern as can affect the lifetime of the compressor relay. 
Although no explicit policy for limiting the number of transitions was formulated in the MPC, we put in place the following two mechanisms to reduce it:
\begin{itemize}
 \item each second cycle  of the PWM control signal is horizontally flipped, as depicted in \citefig{fig:flippedPWM}. This allows to reduce from 2 to 1 the number of transitions per cycle;
 \item on or off PWM pulses with duration shorter than 10~second are ignored. For example, if a given PWM period has a duty cycle shorter than 10~s, the cycle is considered as it was fully off, and vice-versa. 
\end{itemize}
Since the PWM and transformations described above, the control trajectory that is finally actuated is an approximation of the original MPC law. This might lead to violate the temperature constraints during the actuation period that, however, can be chosen small enough to make them negligible. 
Eventual prediction and actuation errors are absorbed by the receding horizon formulation and are taken into account in the following actuation period. Another source of error that impacts the optimality of the actuated control law is the non null computation time required to solve the optimization problem. In fact, while the solver is computing the freezer state is left unaltered with respect to the previous receding horizon cycle; as the length of the receding horizon cycle ($d$, 120~s) must be hardly met, the correct timing of the PWM cycle is compromised, especially if the computation time extends for too long. For example, considering the freezer on from the previous cycle, a computation time of 20~s and $\tau_{\text{on}, k}=0~$s, the finally actuated duty cycle would be 0.16 instead of 0.

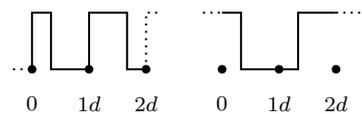
\begin{figure}[!ht]
\centering
{\footnotesize
\begin{tikzpicture}
 \tikzset{>=latex}

\begin{scope}[shift={(0,0)}]
\draw[line width=0.75] (0.0,0.) -- (0.0,0.75) -- (0.25,0.75) -- (0.25,0.0) -- (0.75,0.0) -- 
  (0.75,0.75) -- (1.25,0.75) -- (1.25,0.0) -- (1.50,0.0);
\draw[line width=0.75, dotted] (1.50,0.0) -- (1.50,0.75) -- (1.70,0.75);
\draw[line width=0.75, dotted] (-0.25,0.0) -- (0, 0.);
\filldraw(0.,0.) circle (0.05) node[below=0.25]() {$0$};
\filldraw(0.75,0.) circle (0.05) node[below=0.25]() {$1d$};
\filldraw(1.50,0.) circle (0.05) node[below=0.25]() {$2d$};
\end{scope}

\begin{scope}[shift={(2.5,0)}]
\draw[line width=0.75] (0.0,0.75) -- (0.25,0.75) -- (0.25,0.0) -- (0.75,0.0) -- 
  (0.75,0.0) -- (1.00,0.0) -- (1.0,0.75) -- (1.50,0.75);
\draw[line width=0.75, dotted] (1.50,0.75) -- (1.90,0.75);
\draw[line width=0.75, dotted] (-0.25,0.75) -- (0, 0.75);
\filldraw(0.,0.) circle (0.05) node[below=0.25]() {$0$};
\filldraw(0.75,0.) circle (0.05) node[below=0.25]() {$1d$};
\filldraw(1.50,0.) circle (0.05) node[below=0.25]() {$2d$};
\end{scope}
\end{tikzpicture}
}
\caption{
The PWM signal before (left) and after (right) flipping horizontally the second cycle of the period. This allows to halve the total number of on/off transitions.}\label{fig:flippedPWM}
\end{figure}

\subsection{MPC experimental results}
This section presents the results of a consumption shifting experiment using a MPC-controlled domestic freezer. The objective of the experiment is to assess in practice the performance of the power-to-temperature freezer prediction models, thereby addressing the model selection process also from the application perspective. We compare the performance of three different predictions models, namely Model~B, Model~C and Model~E. The three resulting MPC setups are respectively referred to as MPC-B, MPC-C and MPC-E. 
The sample time $d$ is 120~s, and $N$ is 270, that corresponds to an optimization horizon length of 5~hour. Each MPC experiment lasts for 6~hours. As for the identification experiments, consumer behavior is not considered, \ie\ the freezer content stays unchanged and the door is closed. To assure equal conditions during the experiments, the room temperature was regulated to \SI{23}{\celsius}.



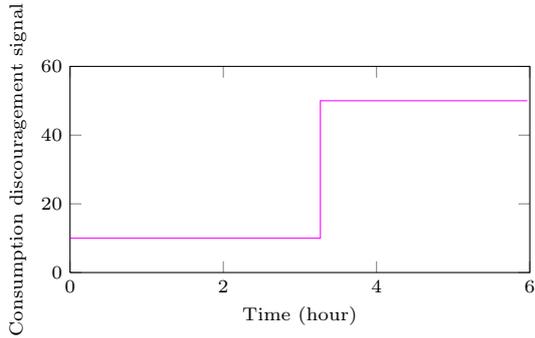
\begin{figure}[!ht]
\centering
\scriptsize
%
%
%
\definecolor{mycolor1}{rgb}{1.00000,0.00000,1.00000}%
\begin{tikzpicture}

\begin{axis}[%
width=\matlabfigurewidth,
height=0.45184331797235\matlabfigurewidth,
scale only axis,
xmin=0,
xmax=6,
xlabel={Time (hour)},
ymin=0,
ymax=60,
ylabel={Consumption discouragement signal}
]
\addplot[const plot,color=mycolor1,solid] plot table[row sep=crcr] {%
0	10\\
0.0333333333333333	10\\
0.0666666666666667	10\\
0.1	10\\
0.133333333333333	10\\
0.166666666666667	10\\
0.2	10\\
0.233333333333333	10\\
0.266666666666667	10\\
0.3	10\\
0.333333333333333	10\\
0.366666666666667	10\\
0.4	10\\
0.433333333333333	10\\
0.466666666666667	10\\
0.5	10\\
0.533333333333333	10\\
0.566666666666667	10\\
0.6	10\\
0.633333333333333	10\\
0.666666666666667	10\\
0.7	10\\
0.733333333333333	10\\
0.766666666666667	10\\
0.8	10\\
0.833333333333333	10\\
0.866666666666667	10\\
0.9	10\\
0.933333333333333	10\\
0.966666666666667	10\\
1	10\\
1.03333333333333	10\\
1.06666666666667	10\\
1.1	10\\
1.13333333333333	10\\
1.16666666666667	10\\
1.2	10\\
1.23333333333333	10\\
1.26666666666667	10\\
1.3	10\\
1.33333333333333	10\\
1.36666666666667	10\\
1.4	10\\
1.43333333333333	10\\
1.46666666666667	10\\
1.5	10\\
1.53333333333333	10\\
1.56666666666667	10\\
1.6	10\\
1.63333333333333	10\\
1.66666666666667	10\\
1.7	10\\
1.73333333333333	10\\
1.76666666666667	10\\
1.8	10\\
1.83333333333333	10\\
1.86666666666667	10\\
1.9	10\\
1.93333333333333	10\\
1.96666666666667	10\\
2	10\\
2.03333333333333	10\\
2.06666666666667	10\\
2.1	10\\
2.13333333333333	10\\
2.16666666666667	10\\
2.2	10\\
2.23333333333333	10\\
2.26666666666667	10\\
2.3	10\\
2.33333333333333	10\\
2.36666666666667	10\\
2.4	10\\
2.43333333333333	10\\
2.46666666666667	10\\
2.5	10\\
2.53333333333333	10\\
2.56666666666667	10\\
2.6	10\\
2.63333333333333	10\\
2.66666666666667	10\\
2.7	10\\
2.73333333333333	10\\
2.76666666666667	10\\
2.8	10\\
2.83333333333333	10\\
2.86666666666667	10\\
2.9	10\\
2.93333333333333	10\\
2.96666666666667	10\\
3	10\\
3.03333333333333	10\\
3.06666666666667	10\\
3.1	10\\
3.13333333333333	10\\
3.16666666666667	10\\
3.2	10\\
3.23333333333333	10\\
3.26666666666667	50\\
3.3	50\\
3.33333333333333	50\\
3.36666666666667	50\\
3.4	50\\
3.43333333333333	50\\
3.46666666666667	50\\
3.5	50\\
3.53333333333333	50\\
3.56666666666667	50\\
3.6	50\\
3.63333333333333	50\\
3.66666666666667	50\\
3.7	50\\
3.73333333333333	50\\
3.76666666666667	50\\
3.8	50\\
3.83333333333333	50\\
3.86666666666667	50\\
3.9	50\\
3.93333333333333	50\\
3.96666666666667	50\\
4	50\\
4.03333333333333	50\\
4.06666666666667	50\\
4.1	50\\
4.13333333333333	50\\
4.16666666666667	50\\
4.2	50\\
4.23333333333333	50\\
4.26666666666667	50\\
4.3	50\\
4.33333333333333	50\\
4.36666666666667	50\\
4.4	50\\
4.43333333333333	50\\
4.46666666666667	50\\
4.5	50\\
4.53333333333333	50\\
4.56666666666667	50\\
4.6	50\\
4.63333333333333	50\\
4.66666666666667	50\\
4.7	50\\
4.73333333333333	50\\
4.76666666666667	50\\
4.8	50\\
4.83333333333333	50\\
4.86666666666667	50\\
4.9	50\\
4.93333333333333	50\\
4.96666666666667	50\\
5	50\\
5.03333333333333	50\\
5.06666666666667	50\\
5.1	50\\
5.13333333333333	50\\
5.16666666666667	50\\
5.2	50\\
5.23333333333333	50\\
5.26666666666667	50\\
5.3	50\\
5.33333333333333	50\\
5.36666666666667	50\\
5.4	50\\
5.43333333333333	50\\
5.46666666666667	50\\
5.5	50\\
5.53333333333333	50\\
5.56666666666667	50\\
5.6	50\\
5.63333333333333	50\\
5.66666666666667	50\\
5.7	50\\
5.73333333333333	50\\
5.76666666666667	50\\
5.8	50\\
5.83333333333333	50\\
5.86666666666667	50\\
5.9	50\\
5.93333333333333	50\\
5.96666666666667	50\\
};
\end{axis}
\end{tikzpicture}%
\caption{The virtual electricity price.}\label{fig:MPC:price}
\end{figure}

The virtual electricity price (implemented in the MPC cost function by the sequence $c_k,\ k=i,\dots,i+N$ ) is shown in \citefig{fig:MPC:price}. 
Considering the MPC formulation (\ie\, penalty function linear in the consumption and freezer temperature allowed in a given range), the expected behavior is that the freezer will reach the lowest allowed temperature before the larger value of $c_k$ in order to decrease as much as possible the consumption of expensive electricity. 
\citefig{fig:MPC:actuatedpower} compares the receding horizon power consumption trajectory as determined by the MPC and the actuated freezer power consumption averaged on a 120 second interval. 
The difference between the two profiles is due to the fact that the freezer consumption is, as mentioned in \citesec{sec:mpcgeneralformulation}, modelled as a constant term, while in the real case it depends on the absorption of the induction motor. \citefig{fig:MPC:instant} shows the freezer power consumption measurements (at 10~s resolution) and the respective average calculated on a 120 second time interval. In the former profile, the effect of the PWM is evident. 
\citefig{fig:MPC:temperatures} shows the freezer temperature as measured during the experiments. As can be seen from \citefig{}, MPC-C is close to reach the lowest temperature (\SI{-26.53}{\celsius} while the limit is \SI{-27}{\celsius}) just before the release of the large virtual electricity price. This evidence indicates that the controller is able to exploit nearly all the storage capacity allowed by the MPC setup.

\begin{figure}[!ht]
\centering
\scriptsize
%
%
\begin{tikzpicture}

\begin{axis}[%
width=\matlabfigurewidth,
height=0.586117511520737\matlabfigurewidth,
scale only axis,
xmin=0,
xmax=6,
xlabel={Time (hour)},
ymin=0,
ymax=70,
ylabel={Average power consumption (W)},
legend style={at={(0.15,0.13531746031746)},anchor=south west,fill=none,draw=none,legend cell align=left}
]
\addplot [color=black,solid,mark=o,mark options={solid}]
  table[row sep=crcr]{%
0.0152777777777778	19.3333333333333\\
0.115277777777778	34.1666666666667\\
0.215277777777778	54.75\\
0.315277777777778	45.5833333333333\\
0.415277777777778	36.9166666666667\\
0.515277777777778	49.25\\
0.615277777777778	39.3333333333333\\
0.715277777777778	36.25\\
0.815277777777778	38.0833333333333\\
0.915277777777778	68\\
1.01527777777778	65.9166666666667\\
1.11527777777778	65\\
1.21527777777778	64.4166666666667\\
1.31527777777778	64\\
1.41527777777778	63.1666666666667\\
1.51527777777778	62.9166666666667\\
1.61527777777778	62.5833333333333\\
1.71527777777778	62.25\\
1.81527777777778	62\\
1.91527777777778	61.25\\
2.01527777777778	61\\
2.11527777777778	60.75\\
2.21527777777778	60.5833333333333\\
2.31527777777778	60\\
2.41527777777778	59\\
2.51527777777778	59.5833333333333\\
2.61527777777778	59.3333333333333\\
2.71527777777778	58.75\\
2.81527777777778	59\\
2.91527777777778	59.25\\
3.01527777777778	59\\
3.11527777777778	58.6666666666667\\
3.21527777777778	58.5833333333333\\
3.31527777777778	0\\
3.41527777777778	0\\
3.51527777777778	0\\
3.61527777777778	0\\
3.71527777777778	0\\
3.81527777777778	0\\
3.91527777777778	0\\
4.01527777777778	0\\
4.11527777777778	0\\
4.21527777777778	0\\
4.31527777777778	37.3333333333333\\
4.41527777777778	68.1666666666667\\
4.51527777777778	66.1666666666667\\
4.61527777777778	35\\
4.71527777777778	0\\
4.81527777777778	38.6666666666667\\
4.91527777777778	39.25\\
5.01527777777778	51.1666666666667\\
5.11527777777778	46\\
5.21527777777778	43.25\\
5.31527777777778	29.5833333333333\\
5.41527777777778	20\\
5.51527777777778	42.25\\
5.61527777777778	18.6666666666667\\
5.71527777777778	48.1666666666667\\
5.81527777777778	25.8333333333333\\
5.91527777777778	37.0833333333333\\
6	0\\
};
\addlegendentry{Actuated};

\addplot [color=red,solid]
  table[row sep=crcr]{%
0	16.4015370272909\\
0.0333333333333333	2.17839351535076e-06\\
0.0666666666666667	25.1383834873991\\
0.1	27.4237774622934\\
0.133333333333333	40.6625519044883\\
0.166666666666667	59.2310759255379\\
0.2	54.4955269144807\\
0.233333333333333	39.1740775208023\\
0.266666666666667	64.1394444379371\\
0.3	46.2073234018576\\
0.333333333333333	47.8060709822148\\
0.366666666666667	47.3697141105818\\
0.4	36.0140189289414\\
0.433333333333333	41.8153966888426\\
0.466666666666667	11.9858543865685\\
0.5	43.7318467038294\\
0.533333333333333	25.1710701925408\\
0.566666666666667	13.7859237662665\\
0.6	37.9115047401715\\
0.633333333333333	14.0316756747784\\
0.666666666666667	43.6355311861353\\
0.7	32.8624799509789\\
0.733333333333333	49.3627900856136\\
0.766666666666667	41.4535839256923\\
0.8	38.7928810764615\\
0.833333333333333	24.3632920857649\\
0.866666666666667	67.9999669352578\\
0.9	65.9265632684296\\
0.933333333333333	67.9999999955871\\
0.966666666666667	67.99999993865\\
1	67.9999998121139\\
1.03333333333333	67.999999999995\\
1.06666666666667	67.9999787881507\\
1.1	67.9999998798594\\
1.13333333333333	67.9999999998963\\
1.16666666666667	67.9999999999914\\
1.2	67.9999999999995\\
1.23333333333333	67.9999999999995\\
1.26666666666667	67.9999999994525\\
1.3	67.9999999999995\\
1.33333333333333	67.9999999991041\\
1.36666666666667	67.9999999935721\\
1.4	67.9999999801466\\
1.43333333333333	67.9999999999909\\
1.46666666666667	67.9999999971874\\
1.5	68\\
1.53333333333333	67.9999999830757\\
1.56666666666667	67.9999999745551\\
1.6	67.9999999999645\\
1.63333333333333	67.9999999999973\\
1.66666666666667	67.9999999976144\\
1.7	67.9999998941844\\
1.73333333333333	67.9999999999782\\
1.76666666666667	67.9999999999882\\
1.8	67.9999999999945\\
1.83333333333333	67.9999999998504\\
1.86666666666667	67.9999999999891\\
1.9	67.9999999294005\\
1.93333333333333	67.9999999983033\\
1.96666666666667	68.0000000000005\\
2	67.9999999447036\\
2.03333333333333	67.9999999814308\\
2.06666666666667	67.9999999835627\\
2.1	67.9999999999986\\
2.13333333333333	67.9999999999668\\
2.16666666666667	67.9999994713553\\
2.2	67.9999999989614\\
2.23333333333333	67.9999999981565\\
2.26666666666667	67.9999999999877\\
2.3	67.9999999992783\\
2.33333333333333	67.9999999814163\\
2.36666666666667	67.9999999995744\\
2.4	67.9999999793436\\
2.43333333333333	67.9999999839029\\
2.46666666666667	67.999999994473\\
2.5	67.9999999999745\\
2.53333333333333	67.9999999117722\\
2.56666666666667	67.9999999999973\\
2.6	67.9999999999918\\
2.63333333333333	67.9999999998709\\
2.66666666666667	67.9999999659381\\
2.7	67.9999999998672\\
2.73333333333333	67.9999999752804\\
2.76666666666667	67.9999999999964\\
2.8	67.9999999999927\\
2.83333333333333	67.9999999999891\\
2.86666666666667	67.9999999989168\\
2.9	67.9999999937859\\
2.93333333333333	67.9999999999482\\
2.96666666666667	67.9999999113948\\
3	68\\
3.03333333333333	68\\
3.06666666666667	67.9999999998752\\
3.1	67.9999999999474\\
3.13333333333333	67.9999999795516\\
3.16666666666667	67.9999999999857\\
3.2	67.9999999828649\\
3.23333333333333	67.9999999998022\\
3.26666666666667	8.1536200013943e-10\\
3.3	-4.54747350886464e-12\\
3.33333333333333	3.24007487506606e-10\\
3.36666666666667	8.16157808003481e-10\\
3.4	4.54747350886464e-13\\
3.43333333333333	4.86579665448517e-11\\
3.46666666666667	1.43700162880123e-10\\
3.5	0\\
3.53333333333333	4.57703208667226e-10\\
3.56666666666667	2.29420038522221e-10\\
3.6	8.62200977280736e-10\\
3.63333333333333	2.27373675443232e-13\\
3.66666666666667	2.04636307898909e-11\\
3.7	1.24828147818334e-10\\
3.73333333333333	4.54747350886464e-13\\
3.76666666666667	-2.27373675443232e-13\\
3.8	5.6843418860808e-12\\
3.83333333333333	2.27373675443232e-13\\
3.86666666666667	2.33824948736583e-07\\
3.9	5.6843418860808e-12\\
3.93333333333333	0\\
3.96666666666667	7.15867827238981e-08\\
4	6.80013272358337e-08\\
4.03333333333333	1.15701368486043e-08\\
4.06666666666667	7.12179826223291e-09\\
4.1	4.67068730358733e-08\\
4.13333333333333	7.71610075389617e-07\\
4.16666666666667	2.68300937023014e-11\\
4.2	3.25796918332344e-08\\
4.23333333333333	4.75119894254749\\
4.26666666666667	15.7883115237853\\
4.3	26.295370701162\\
4.33333333333333	43.208482175223\\
4.36666666666667	57.7436643748044\\
4.4	62.8014914139314\\
4.43333333333333	67.9999999904182\\
4.46666666666667	50.9896655501448\\
4.5	65.774632579031\\
4.53333333333333	40.8108079235978\\
4.56666666666667	46.428876997247\\
4.6	29.7239397410192\\
4.63333333333333	31.6028075052061\\
4.66666666666667	24.3464588566783\\
4.7	7.50843695338813\\
4.73333333333333	32.2490208954173\\
4.76666666666667	6.00981914285558\\
4.8	30.5985271349821\\
4.83333333333333	2.88764567812905e-11\\
4.86666666666667	28.6654006522074\\
4.9	32.1814535004492\\
4.93333333333333	25.7258313937195\\
4.96666666666667	27.5542776498035\\
5	52.6190597376749\\
5.03333333333333	45.019126658699\\
5.06666666666667	50.4950632955924\\
5.1	39.1283638606819\\
5.13333333333333	44.1334854400745\\
5.16666666666667	13.6387231561396\\
5.2	44.1970744018563\\
5.23333333333333	6.5116390659939\\
5.26666666666667	49.3492736060496\\
5.3	22.1714476233678\\
5.33333333333333	45.51354563538\\
5.36666666666667	26.242405599148\\
5.4	21.4921298744091\\
5.43333333333333	29.7601852353134\\
5.46666666666667	33.595284943216\\
5.5	37.3693155534913\\
5.53333333333333	28.8901728654109\\
5.56666666666667	33.386771822449\\
5.6	21.7956266857741\\
5.63333333333333	39.1308692674224\\
5.66666666666667	32.1728853366828\\
5.7	41.3224202105746\\
5.73333333333333	35.2150519721604\\
5.76666666666667	38.4522271145888\\
5.8	30.8026283285203\\
5.83333333333333	28.7286730350736\\
5.86666666666667	21.965462493504\\
5.9	30.8553051104004\\
5.93333333333333	30.6059721866768\\
5.96666666666667	36.086059080179\\
};
\addlegendentry{MPC law};

\end{axis}
\end{tikzpicture}%
\caption{The MPC-C power consumption control law and the actuated freezer power consumption averaged on a 2 minute interval.}\label{fig:MPC:actuatedpower}
\end{figure}
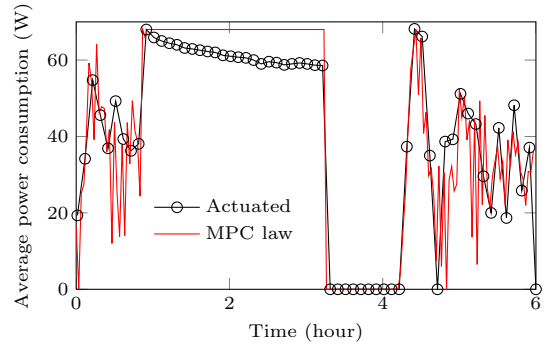

\begin{figure}[!ht]
\centering
\scriptsize
\input{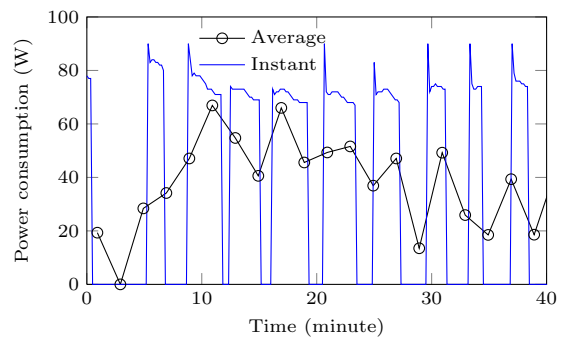}
\caption{The real-time freezer power consumption (sampled at 10~s) and the average on a 2 minute interval during a portion of the MPC-C experiment.}\label{fig:MPC:instant}
\end{figure}

\begin{figure}[!ht]
\centering
\scriptsize
\input{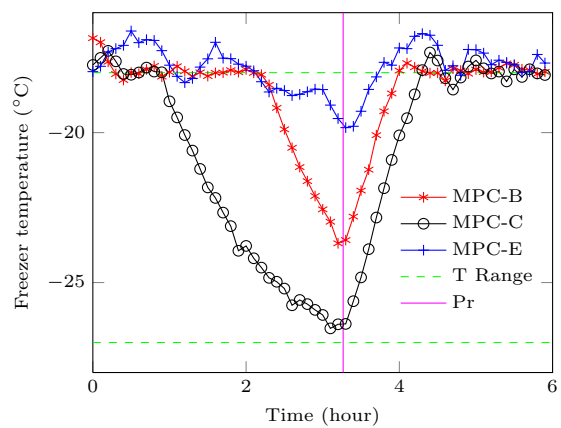}
\caption{The freezer temperature during the three MPC experiments. The horizontal dashed line denotes the temperature constraints, while the continuous vertical line indicates the instant of time when the large value of the virtual electricity price is released.}\label{fig:MPC:temperatures}
\end{figure}

In order to formally compare the performance of the different MPC setups, we use the following three metrics:
\begin{itemize}
 \item[$m_0$:] the value of the MPC cost function \eqref{eq:MPC:realcost} evaluated, for each experiment, considering the actuated consumption profile and measured temperature.
 \item[$m_1$:] amount of electricity shifted (or virtual storage capacity), evaluated as the nominal power of the freezer times the duration between the instant of time when the larger value of the virtual electricity price is triggered until when the freezer temperature reaches the upper threshold;
 \item[$m_2$:] distance from the upper temperature limit when the upper bound constraint is violated on the duration of the experiments in number of discrete time steps $L$. Formally it is as:
 \begin{align}
  m_2 = \sum \frac{1}{L} v_k
  \end{align}
  where
  \begin{align} 
  v_k = \begin{cases}
  0, & T_k <= T_\text{max}\\
  T_k - T_\text{max}, & T_k > T_\text{max} 
  \end{cases},
  \end{align}
  for $k=0, \dots, L-1$;
  \item[$m_3$:] maximum violation of the temperature upper bound constraint ($T_\text{max}$).
\end{itemize}
The metrics calculated for the MPC experiments are summarized in \citetable{tab:mpcs:metrics}. The metrics $m_0$ and $m_1$ are in favor of MPC-C, meaning that it achieves the lowest cost of operation and largest shift of consumption.
As far as the temperature comfort metrics are concerned, the MPC-B achieves the smallest violation of the upper temperature bound constraint ($m_3$), although MPC-C has the lowest average violation ($m_2$). 
Nonlinear Model~E, that had the best prediction performance in the previous section, does not perform well in the respective MPC setup. 
This is to ascribe to the fact that the actuated control law is strongly suboptimal. Even if nonconvex optimization algorithm for global optimum search are available (like PSO or genetic algorithms), they have been not considered because the associated computational burden is relevant. Moreover, considering the experimental results, it can be said that implementing complex nonconvex algorithm is not justified by the performance of the MPC-C that, as discussed while describing \citefig{fig:MPC:temperatures}, is already able to exploit nearly all the flexibility of the freezer while showing overall a good capability of satisfying freezer constraints. Another solution, that however has not been attempted in the proposed experiments for the same reason explained above, could consists in formulating a convex optimization problem by linearizing nonlinear Model~E as similarly done for the state estimation in the extended Kalman filter.



\begin{table}[!ht]
\centering
{ \small
\renewcommand{\arraystretch}{1.5}
\caption{Summary of MPC Performance metrics.}\label{tab:mpcs:metrics}
\begin{tabular}{C{1.90cm} C{1.5cm} C{0.75cm} C{0.75cm} C{0.75cm}}
\hline
Implemented Model & $m_0$ \ \ \ (cost) & $m_1$ (Wh) & $m_2$ (\SI{}{\celsius})& $m_3$ (\SI{}{\celsius})\\
\hline
MPC-B & \SI{1.7e5} & 52.1 & 0.06 & 0.43 \\
MPC-C & \SI{1.5e5} & 74.8 & 0.04 & 0.68 \\
MPC-E & \SI{2.0e5} & 38.5 & 0.39 & 1.29 \\
\hline
\end{tabular}
}
\end{table}

In addition to the proposed metrics, \citetable{tab:mpcs:other} reports the average computation time required to determine the MPC solution (single thread process on an Intel i7 2.10~GHz) and the round trip storage efficiency, calculated as the ratio between the amount of electricity invested to reach the freezer lower temperature and the one that is harvested.  As expected, the computation time for linear models is lower than for the nonlinear case, that requires to solve a nonconvex optimization problem. The round trip efficiency of MPC-C is the lowest because, by achieving a lower temperature, thermal losses are increased.

\begin{table}[!ht]
\centering
{ \small
\renewcommand{\arraystretch}{1.5}
\caption{Additional performance metrics.}\label{tab:mpcs:other}
\begin{tabular}{C{2cm} C{1.9cm} C{1.8cm}}
\hline
Implemented Model & Computation time (s) & Round trip efficiency \\
\hline
MPC-B & 4.5 & 82\%  \\
MPC-C & 5.7 &  51\% \\
MPC-E & 21.6 & 54\% \\
\hline
\end{tabular}
}
\end{table}

\section{Conclusions and perspectives}
We described the application of grey-box modeling to identify suitable power consumption-to-temperature models of a domestic refrigeration using experimental measurements from an instrumented 333~liter freezer. Consumer behavior was not considered at this stage. Models were formulated using stochastic differential equations (SDEs) and identified used maximum likelihood estimation (MLE). The proposed models were validated by checking the model residuals correlation, and model extensions were cross-validated to detect model over-fitting. Among the presented linear models, it has been shown that a third order model is able to capture nearly all the dynamics contained in the measurements. While the extension to a fourth order linear model was not justified by statistical evidence, it was shown that implementing a nonlinear description of the reverse Carnot cycle leads to a marginal improvement of the model prediction performance.
As the modeling effort was framed within the context of intelligent energy strategies for demand side management, the second part of the paper is devoted to assessing the models performance in a demand response experiment. It consisted in quantifying the virtual storage action that a freezer can achieve using model predictive control (MPC) and the prediction models previously identified.
From the experiments, it emerged that the third order linear model was able to harvest nearly all the flexibility inherent the freezer operation. Also, it was seen that the mathematical formulation plays in favor of linear models because they result in convex optimization problems that are tractable and efficient to solve.
In the best performing experiment, a virtual storage capacity equivalent to 75~Wh of electrical energy with a round trip efficiency of 51\% was measured. Although consumer behavior was not considered in the experiments, these figures already give an overview on the amount of storage capacity that is possible to harvest from the freezer operation in an ideal condition. Considering the low specific energy density, further studies should be devoted to asses the cost-benefit of the potential deployment of freezer-based demand response programs, therefore with an economic assessment of the cost for the hardware necessary to achieve flexible operation. Phase change materials (PCMs), which have been proven to enhance storage capacity \cite{taneja2013impact}, should be indeed considered in future identification and MPC experiments. Another challenging aspect regards the identification of consumer behavior (in terms of both additional food load, that might contribute to increase storage performance, and door opening that could cause a quick drop of achieved storage level), that is relevant for estimating the freezer flexibility in a real operating scenario. In this context, the proposed grey-box freezer models could be used as a backbone where to plug models of the consumer behavior. The latter kind of models might rely on completely different modeling approaches (\eg\ pattern recognition techniques) and can use the former as an interface for the physical heat transfer principles. Overall, this strategy would result in a model-based approach and could be compared with model-free techniques (such as in \cite{DBLP:journals/corr/RuelensICB15}).

\appendix

\section{Identified Model Parameters}\label{app:modelpars}
Tables A.6-A.10 shows the identified model parameters.

\begin{table}[!ht]
\centering
{ \scriptsize
\renewcommand{\arraystretch}{1.0}
\caption{Model~A identification summary.}\label{tab:modelA:pars}
\begin{tabular}{l c l l}
\hline
Parameter & Unit & Value & $\sigma$ \\
\hline
$C_a$ & \si{\joule\per\kelvin} & \SI{2.99e+3}{} & \SI{5.00e+4}{}\\
$R_w$ & \si{\kelvin\per\watt} & \SI{5.69e-1}{} & \SI{7.50e-1}{}\\
$\alpha$ & -- &\SI{-4.54} & \SI{7.0e-3}{}\\
\COP & -- & \SI{3.01e-1}{} & \SI{6.07e-3}{}\\
\multicolumn{2}{l}{log-likelihood value} & 19833.1 & -\\
\hline
\end{tabular}
}
\end{table}

\begin{table}[!ht]
\centering
{ \scriptsize
\renewcommand{\arraystretch}{1.0}
\caption{Model~B identification summary.}\label{tab:modelB:pars}
\begin{tabular}{l c l l}
\hline
Parameter & Unit & Value & $\sigma$ \\
\hline
$C_a$ & \si{\joule\per\kelvin} & \SI{9.74e+3}{} & \SI{2.12e+3}{}\\
$C_e$ &\si{\joule\per\kelvin}  & \SI{2.28e+3}{} & \SI{1.67e+2}{}\\
$R_e$ &\si{\kelvin\per\watt}   & \SI{9.74e-2}{} & \SI{7.03e-3}{}\\
$R_w$ &\si{\kelvin\per\watt}   & \SI{9.93e-1}{} & \SI{1.80e-1}{}\\
$\alpha_0$ & -- & \SI{-1.36e+1}{} & \SI{2.76}{} \\
$\alpha_1$ & -- & \SI{-2.59}{}  & \SI{1.45e-01}{} \\
\COP & -- & \SI{1.04e+0}{} & \SI{1.87e-1}{} \\
\multicolumn{2}{l}{log-likelihood value}  & 25165.4 & -- \\
\hline
\end{tabular}
}
\end{table}

\begin{table}[!ht]
\centering
{ \scriptsize
\renewcommand{\arraystretch}{1.0}
\caption{Model~C identification summary.}\label{tab:modelC:pars}
\begin{tabular}{l c l l}
\hline
Parameter & Unit & Value & $\sigma$ \\
\hline
$C_a$ &\si{\joule\per\kelvin} & \SI{4.76e+3}{} & \SI{2.45e+3}{}\\
$C_e$ &\si{\joule\per\kelvin} & \SI{1.05e+3}{} & \SI{8.37e+2}{}\\
$C_w$ &\si{\joule\per\kelvin} & \SI{8.11e+3}{} & \SI{4.60e+3}{}\\

$R_a$ &\si{\kelvin\per\watt} & \SI{4.97e-1}{} & \SI{2.82e-1}{}\\
$R_e$ &\si{\kelvin\per\watt} & \SI{1.12e-1}{} & \SI{8.14e-2}{}\\
$R_w$ &\si{\kelvin\per\watt} & \SI{1.28e-0}{} & \SI{7.20e-1}{}\\

$\alpha_0$ & -- & \SI{-8.31}{} & \SI{1.7e-02}{} \\
$\alpha_1$ & -- & \SI{-6.83}{} & \SI{3.8e-02}{}\\
$\alpha_2$ & -- & \SI{-3.15}{} & \SI{2.6e-01}{}\\
\COP & -- &\SI{7.68e-1}{} & \SI{1.65e-1}{}\\
\multicolumn{2}{l}{log-likelihood value} & 25168.9 & -- \\
\hline
\end{tabular}
}
\end{table}

\begin{table}[!ht]
\centering
{ \scriptsize
\renewcommand{\arraystretch}{1.0}
\caption{Model~D identification summary.}\label{tab:modelD:pars}
\begin{tabular}{l c l l}
\hline
Parameter & Unit & Value & $\sigma$ \\
\hline
$C_a$ & \si{\joule\per\kelvin} & \SI{1.25e+04}{} & \SI{2.02e+3}{}\\
$C_e$ & \si{\joule\per\kelvin} & \SI{1.22e+03}{} & \SI{2.63e+3}{}\\
$C_w$ & \si{\joule\per\kelvin} & \SI{5.23e+03}{} & \SI{1.63e+3}{}\\
$C_f$ & \si{\joule\per\kelvin} & \SI{3.94e+03}{} & \SI{8.60e+4}{}\\

$R_a$ & \si{\kelvin\per\watt}  & \SI{4.81e-01}{} & \SI{1.17e-5}{}\\
$R_e$ & \si{\kelvin\per\watt}  & \SI{1.12e-01}{} & \SI{1.90e-1}{}\\
$R_w$ & \si{\kelvin\per\watt}  & \SI{6.25e-01}{} & \SI{9.34e-2}{}\\
$R_f$ & \si{\kelvin\per\watt}  & \SI{1.08e00}{} & \SI{1.92e-1}{}\\

$\eta$ & -- & \SI{5.38e-1}{} & \SI{2.97e-1}{}\\
$\alpha_0$ & -- & -7.20 &  \SI{1.1e-1}{}\\
$\alpha_1$ & -- & -3.66 & \SI{3.4e-2}{}\\
$\alpha_2$ & -- & \SI{-1.1e1}{} & \SI{2.2e-2}{}\\
\multicolumn{2}{l}{log-likelihood value} & 25169.9 & -- \\
\hline
\end{tabular}
}
\end{table}

\begin{table}[!ht]
\centering
{ \scriptsize
\renewcommand{\arraystretch}{1.0}
\caption{Model~E identification summary.}\label{tab:modelE:pars}
\begin{tabular}{l c l l}
\hline
Parameter & Unit & Value & $\sigma$ \\
\hline
$C_a$ & \si{\joule\per\kelvin} & \SI{1.25e+04}{} & \SI{3.80e-1}{}\\
$C_e$ & \si{\joule\per\kelvin} & \SI{1.22e+03}{} & \SI{1.03e+2}{}\\
$C_w$ & \si{\joule\per\kelvin} & \SI{8.30e+03}{} & \SI{1.59e+2}{}\\
$R_a$ & \si{\kelvin\per\watt}  & \SI{1.61e-01}{} & \SI{8.66e-5}{}\\
$R_e$ & \si{\kelvin\per\watt}  & \SI{1.47e-01}{} & \SI{8.49e-4}{}\\
$R_w$ & \si{\kelvin\per\watt}  & \SI{6.32e-01}{} & \SI{3.53e-3}{}\\
$\eta$ & -- & \SI{5.67e-1}{} & \SI{2.97e-1}{}\\
$\alpha_0$ & -- & -7.20 &  \SI{1.1e-1}{}\\
$\alpha_1$ & -- & -3.66 & \SI{3.4e-2}{}\\
$\alpha_2$ & -- & \SI{-1.1e1}{} & \SI{2.2e-2}{}\\
\multicolumn{2}{l}{log-likelihood value} & 25187.6 & -- \\
\hline
\end{tabular}
}
\end{table}


\section{Formulation of the MPC optimization problem}\label{app:mpcdev}
\subsection*{Freezer linear models}
The discretized version of the linear continuous time stochastic state space model \eqref{eq:ssd:1} can be expressed as:
\begin{align}
 & \boldsymbol{x}_{i+1} = A_d \boldsymbol{x}_i + B_d \boldsymbol{u}_i + W\omega_i\label{eq:mpc:derivation00},
\end{align}
with
\begin{align}
 & A_d = Ad + \mathbb{I}_{n\times n} \\
 & B_d = Bd
\end{align}
where $d$ is the sampling time and the other symbols are as defined in \citesec{methods:model_formulation}. The observation equation is as in \eqref{eq:ssd:2}.
For the moment, we shall assume that $\boldsymbol{u}=P$ and $B \in \mathbb{R}^n$. We will extend to the multi-input case later.
The \emph{expected value} of the vector state for $i=0$ is as:
\begin{align}
 & \overline{\boldsymbol{x}}_1 = A_d \overline{\boldsymbol{x}}_0 + B_d P_0 \label{eq:mpc:derivation0}.
\end{align}
Therefore, the evolution for $i=2$ is as:
\begin{align}
 & \overline{T}_2 = C\big( A_d \overline{\boldsymbol{x}}_1 + B_d P_{1} \big)\label{eq:mpc:derivation1}.
\end{align}
Replacing the second last expression in the last yields to:
\begin{align}
 \overline{T}_2 = &CA_d \big( A_d \overline{\boldsymbol{x}}_0 + B_d P_0 \big) + CB_d P_{1}.
\end{align}
Iterating until the time instant $N$ finally gives:
\begin{align}
\overline{T}_N = &CA_d^N \boldsymbol{x}_0 + CA_d^{N-1}B_d P_{0} + \cdots + CB_dP_{N-1}.
\end{align}
Denoting the sequences $[\overline{T}_1, \dots, \overline{T}_N]^T$, $[{P}_0, \dots, P_{N-1}]^T$ with $\boldsymbol{T}$ and $\boldsymbol{P}$ respectively, the expression above can be reformulated using the matrix product notation as:
\begin{align}
 \boldsymbol{T} = \Phi \overline{\boldsymbol{x}}_0 +\Theta(B) \boldsymbol{P} \label{eq:mpc:compactoutput},
\end{align}
where
\begingroup
\setlength{\arraycolsep}{3pt}
\begin{align}
  \Phi &= \begin{bmatrix}
  CA_d &
  \cdots &
  CA_d^{N}
\end{bmatrix}^T\\
 \Theta(B) &= \begin{bmatrix}
  CA_dB_d & 0 & \cdots & 0\\
  \vdots & \vdots & \ddots & \vdots\\
  CA_d^{N}B_d & CA_d^{N-1}B_d & \cdots & CB_d
\end{bmatrix}.
\end{align}
\endgroup
In the multiple input case, for example as in the case with freezer where $B_d \in \mathbb{R}^{n\times 2} = [B_{d, 0},~B_{d, 1}]$ and $\boldsymbol{u}^T=[P,~T_r]$, the expected value of the state vector is as:
\begin{align}
& \overline{\boldsymbol{x}}_{i+1} = A_d \overline{\boldsymbol{x}}_i + B_{d,0} P_i + B_{d,1} T_{r,i}
\end{align}
and the system output can be therefore expressed as:
\begin{align}
 \boldsymbol{T} = \Phi \overline{\boldsymbol{x}}_0 +\Theta(B) \boldsymbol{P} + \Theta(E) \boldsymbol{v} \label{eq:mpc:compactfinal},
\end{align}
\ie adding the input and the respective transition matrix $\Theta(E)$. \citeeq{eq:mpc:compactfinal} can be used to express the inequality constraints in \eqref{eq:Tconstr}. Finally, the optimization problem \eqref{eq:MPC:realcost}-\eqref{eq:MPC:reale} can be expressed in a standard convex form as:
\begin{align}
 \{ \boldsymbol{P}^o, \boldsymbol{s}^o\} = \argmin{\{ \boldsymbol{P}, \boldsymbol{s}\} \in \Omega}{ \left( \boldsymbol{P}^T \boldsymbol{c} + \boldsymbol{s}^T \boldsymbol{c} \right)}
\end{align}
subject to:
\begin{align}
\Phi \overline{\boldsymbol{x}}_0 +\Theta(B) \boldsymbol{P} + \Theta(E) \boldsymbol{v} &\leq T_\text{max} \\
-\Phi \overline{\boldsymbol{x}}_0 -\Theta(B) \boldsymbol{P} - \Theta(E) \boldsymbol{v} &\leq -T_\text{min} \\
\boldsymbol{P} &\leq \boldsymbol{P}_\text{max} \\
-\boldsymbol{P} &\leq 0  \\
-\boldsymbol{s} &\leq 0
\end{align}
that can be solved for example using the \emph{linprog} function in Matlab.

\subsection*{Freezer nonlinear model}
The discretized version of the nonlinear model is obtained in the same way as for the linear models. However, in this case the coefficient $B_d$ will depend on the state variable. When iterating the time expansion as previously done to determine \eqref{eq:mpc:compactoutput}, the temperature evolution is not a linear function of the decision vector $\boldsymbol{P}$ anymore. This leads to a nonconvex optimization problem.  The final problem is formulated as in \eqref{eq:MPC:realcost}-\eqref{eq:MPC:reale} and solved using the Matlab function \emph{fmincon}.

 \section*{Acknowledgements}
 We would like to thank the anonymous reviewers for their valuable comments and the Editor-in-Chief of \emph{Sustainable Energy, Grids and Networks} (SEGAN), Mr. Mario Paolone, for having extended the revision deadline, allowing us to accomplish the final version of this work.
 
\section*{References}
\bibliography{biblio}

\end{document}